# Core@Shell AgBr@CsPbBr$_3$ Nanocrystals as Precursors to Hollow Lead Halide Perovskite Nanocubes

Zhanzhao Li[1], Yurii P. Ivanov[2], Anna Cabona[1,3], Andrea Fratelli[1,4], Stefano Toso,[1] Saptarshi Chakraborty[4], Giorgio Divitini[2]*, Ilka Kriegel[3], Sergio Brovelli[4]*, Liberato Manna[1]*

[1] Nanochemistry, Istituto Italiano di Tecnologia, Via Morego 30, 16163, Genova, Italy

[2] Electron Spectroscopy and Nanoscopy, Istituto Italiano di Tecnologia, Via Morego 30, 16163, Genova, Italy

[3] Department of Applied Science and Technology, Politecnico di Torino, Corso Duca degli Abruzzi 34, 10129, Turin, Italy

[4] Dipartimento di Scienza dei Materiali, Università degli Studi di Milano Bicocca, via R. Cozzi 55, 20125, Milano, Italy

**ABSTRACT:** We report the synthesis of colloidal core@shell AgBr@CsPbBr$_3$ nanocubes by a one-pot approach, where the nucleation and growth of AgBr nanocrystals occurs rapidly after the injection of chemical precursors. This is immediately followed by the overgrowth of CsPbBr$_3$, delivering AgBr@CsPbBr$_3$ nanocubes of several tens of nanometers in size, with the volume of the AgBr core being only a small fraction of the overall nanocrystal volume. The formation of a core@shell geometry is facilitated by the epitaxial compatibility between AgBr and CsPbBr$_3$ along multiple crystallographic directions. Exchange with Cl$^-$ ions leads to Ag@CsPbCl$_3$ nanocubes, whereas exchange with I$^-$ ions leads to hollow CsPbI$_3$ nanocubes, due to selective etching of the AgBr (or Ag) core region by the I$^-$ ions diffusing in the nanocubes. These hollow CsPbI$_3$ nanocubes can then be converted into hollow CsPbBr$_3$ and CsPbCl$_3$ nanocubes by halide exchange. The optical emission properties of the hollow CsPbX$_3$ (X=Cl, Br, I) nanocubes are in line with those expected from large, non-hollow halide perovskite nanocrystals, indicating that the small hollow region in the cubes has no major influence on their optical properties.

INTRODUCTION

Lead halide perovskite nanocrystals (NCs) have attracted significant research attention in the past decade due to their appealing optical properties, prompting their investigation in various applications.[1-5] Control over size and shape of these types of NCs has reached a high level of maturity, and this has gone hand in hand with a deeper understanding of the kinetics and thermodynamics of NC growth.[6-8] Yet, compared to NCs of more traditional semiconductors (for example II-VI and III-V), the synthesis of core-shell NCs based on halide perovskites and, more generally, metal halides, has been less successful, with only a handful cases reported to date.[9, 10] The difficulty in matching metal halides with enough similarities in crystal structures and lattice parameters to attain core-shell structures is compounded by two additional factors: i) the often rapid halide inter-diffusion that can quickly alloy initially segregated domains with different halide composition; ii) the lability of metal halide NCs, as they might not withstand the conditions required for a shell growth. Another less explored area of research in metal halide NCs is that of hollow geometries, with only a few routes explored to date. The Kirkendall effect has been used to prepare hollow nanostructures of various materials (metals[11], metal oxides[12], metal chalcogenides and phosphides[13, 14]) and has been recently extended to prepare hollow halide perovskites NCs.[15, 16] On the other hand, the high ionic diffusivity in halide perovskites,[17, 18] coupled with the ease of dissolution of metal halides under various stimuli,[19, 20] might provide novel routes to generate hollow geometries in NCs of these materials in addition to the Kirkendall effect.

In this work, we have developed a one-pot synthesis of core@shell AgBr@CsPbBr$_3$ nanocubes. The approach is based on a modification of a standard synthesis protocol for CsPbBr$_3$ NCs with the addition of Ag$^+$ ions in the reaction environment, along with the precursors needed to grow the CsPbBr$_3$ NCs.[21] The much lower solubility of AgBr compared to CsPbBr$_3$ under the reaction conditions of our experiments results in the fast nucleation of AgBr seeds as the first event in the synthesis. Such nucleation quickly deprives the reaction environment of Ag$^+$ ions and sets the conditions for the subsequent growth of a thick epitaxial shell of CsPbBr$_3$ around the AgBr seeds, facilitated by the similarity in lattice constants of AgBr and CsPbBr$_3$ (**Scheme 1a,b**), as also ascertained by us using the recently developed Ogre library for the prediction of ionic epitaxial interfaces.[22] The synthesis delivers cuboidal core@shell AgBr@CsPbBr$_3$ NCs of several tens of nm in edge length. We also verified that the additional presence of Zn$^{2+}$ cations in the synthesis leads to AgBr@CsPbBr$_3$ NCs with a narrower size distribution than when Zn$^{2+}$ is absent.

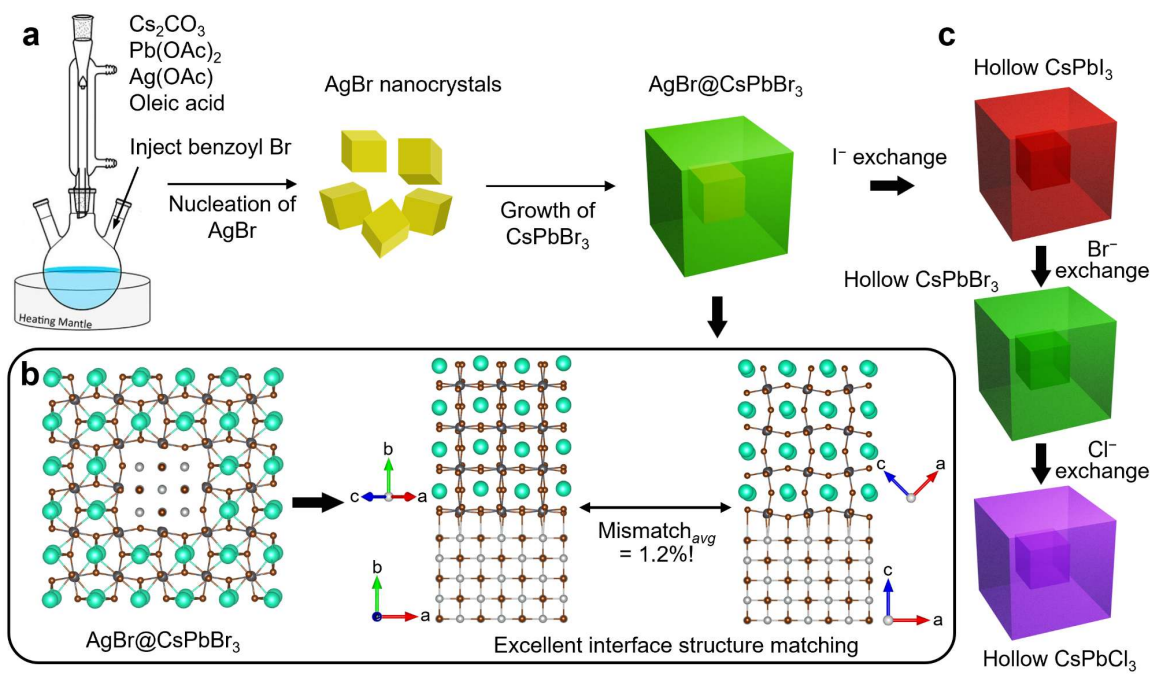

**Scheme 1.** (a) Schematic representation of the synthesis route of core@shell AgBr@CsPbBr$_3$ nanocubes; (b) structural models indicating good lattice matching at the interface between AgBr and CsPbBr$_3$. The models on the right have been prepared using the Ogre library[22]; (c) Hollow CsPbX$_3$ (X=Cl, Br, I) nanocubes obtained by halide exchange reactions. In the first case (hollow CsPbI$_3$ NCs), the Br$^-$ to I$^-$ exchange is accompanied by the dissolution of the core.

Under the transmission electron microscope (TEM) the samples appeared as CsPbBr$_3$ cubes with a small central cavity partially occupied by an Ag-rich domain, which could be either AgBr or metallic Ag due to photodegradation by exposure to ambient light and/or electron irradiation during sample preparation/analysis. These NCs were then subjected to halide exchange reactions. While a complete exchange of Br$^-$ with Cl$^-$ ions led to CsPbCl$_3$ NCs containing a Ag-rich domain inside the cavity, a complete exchange of Br$^-$ with I$^-$ ions led to pure, hollow CsPbI$_3$ NCs, with no remaining Ag inside (**Scheme 1c**). The dissolution of the central AgBr (or Ag) domain in the core@shell NCs was attributed to its reaction with the I$^-$ ions diffusing in the NCs. These hollow CsPbI$_3$ NCs could then be used to generate pure CsPbBr$_3$ and CsPbCl$_3$ hollow cubes by sequential halide exchanges (**Scheme 1c**). Time-resolved PL measurements revealed long lifetimes consistent with low quantum confinement, and transient absorption (TA) measurements evidenced biexciton dynamics in line with the expected volume scaling, suggesting that the photophysics of the hollow cubes is largely determined by their size, with no apparent effect of the inner hollow region.

RESULTS AND DISCUSSION

**Simulations.** We performed simulations of the possible interface structures between AgBr and CsPbBr$_3$, which confirmed excellent structural compatibility between the two phases (see **Figure S1, S2** and **Table S1** of the Supporting Information (SI)), with the lowest interfacial energy being that of the (100)/(100) AgBr/CsPbBr$_3$ configuration. This suggests that a core/shell architecture is feasible and additionally that the two materials might preferentially share these types of low energy interfaces.

**Synthesis, characterization and growth mechanism of the AgBr@CsPbBr$_3$ nanocubes.** The synthesis consists of injecting a solution of benzoyl bromide in a mixture of metal oleate complexes (Cs, Pb, Zn, and Ag oleate) dissolved in excess oleic acid and hexadecane at 100 °C and letting the reaction run for one minute, after which the reaction was quenched by immersing the flask in an ice-water bath. The role of Zn in the synthesis is discussed in detail later. The products of this synthesis, as seen under TEM, were cubes (inset of **Figure 1a**) with 45 ± 7 nm in lateral size. The cubes had a distinct core@shell morphology, with a core region at the center of the cubes that in most cases was a cubic cavity, of around 12 ± 1.8 nm in size, carrying inside an approximately round, higher-contrast domain of 9.9 ± 1.7 nm in size (see **Figure S3** for estimates of sizes of NCs and related core regions). Hence, the overall volume fraction originally occupied by the AgBr core is only around 2% of the whole NC volume. The optical absorption and PL spectra of colloidal suspensions of this sample were compatible with what is expected for large CsPbBr$_3$ NCs, with a tail in the absorption spectrum attributed to the strong light scattering due to partial aggregation of such large NCs in solution (**Figure 1a**).

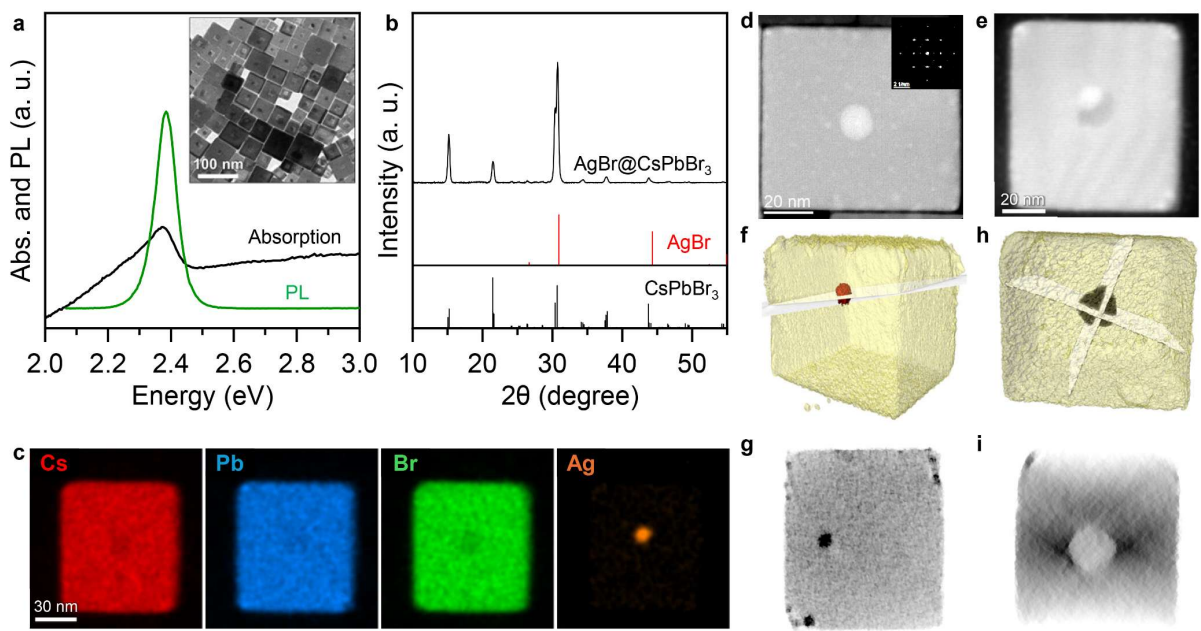

**Figure 1.** Characterization of AgBr@CsPbBr$_3$ NCs. (a) Optical absorption and PL spectra, TEM micrograph (inset) and (b) XRD pattern of a sample of AgBr@CsPbBr$_3$ NCs. In (b), the black reference marks are for CsPbBr$_3$ (ICSD number 97851), the red ones are for AgBr (ICSD number 56546); (c) EDX elemental mapping of a single NC; (d-i) additional microscopy characterizations of two different individual NCs. The left panels (d, f, g) refer to one NC, the right panels (e, h, i) refer instead to another NC and actually the same NC on which EDX analysis is reported in panels (c). (d, e) STEM-HAADF images (inset in d: FFT), (f, h) corresponding 3D renders from STEM-HAADF tomography, and (g, i) ortho-slices from the reconstructed volumes shown in (f) and (h), where the slicing planes are also highlighted.

The X-ray powder diffraction (XRD) pattern could be unambiguously matched to the orthorhombic CsPbBr$_3$ phase (**Figure 1b**), with no apparent presence of other phases, such as metallic Ag or AgBr. However, we note that the detection of AgBr by XRD would be particularly challenging due to the exact overlap with CsPbBr$_3$ peaks and to the small volume ratio of AgBr in the sample. Elemental analysis by inductively coupled plasma optical emission spectroscopy (ICP-OES) confirmed the presence of silver in the sample, with an Ag:Pb atomic ratio of 0.1 (**Table S2**), not far from the estimated Ag:Pb atomic ratio of 0.08 if one considers the ~2% volume fraction initially occupied by the AgBr core. Elemental mapping by energy dispersive X-ray spectroscopy (EDX) of a single NC revealed the presence of a small Ag-rich domain in the cavity (**Figure 1c**).

A more in-depth structural analysis of the NCs was performed by high-angle annular dark-field high-resolution scanning TEM (STEM-HAADF). **Figure 1d** is a STEM-HAADF image of a rare case of core@shell NC in which no void region is observed and the core appears as a bright central region in the image due to the higher mass density of AgBr (or Ag). Fast Fourier transform (FFT) analysis (inset of **Figure 1d**) confirmed that the NC is essentially a monocrystalline CsPbBr$_3$. AgBr has an excellent lattice match with CsPbBr$_3$ (with a mismatch of 1.2%, as shown in **Scheme 1b**), making it challenging to distinguish the AgBr lattice reflections overlapping with those from the thick CsPbBr$_3$ shell. **Figure 1e** is a high resolution STEM image of the much more common case of a NC with a cubic shaped cavity and a Ag-rich domain inside. This is actually the same NC on which EDX mapping is reported in Figure 1c. To identify the three-dimensional structure of these two NCs, we performed STEM-HAADF tomography. **Figure 1f, h** reports the reconstructed volumes of the NCs with ortho-slices (i.e. section cuts through the reconstruction), evidencing a solid core region in one case (**Figure 1g**) and a cavity in the other case (**Figure 1i**). The cubic shape of the cavity is likely dictated by the original morphology of the AgBr domain, which, as indicated by the simulations reported in the SI, would need to adopt a cubic shape to minimize its interface energy with CsPbBr$_3$ through the preferential formation of (100)/(100) interfaces. Yet, we cannot entirely exclude that the initial shape of the AgBr core deviates from a perfect cube and the cubic shape of the cavity arises from a partial reorganization of the CsPbBr$_3$ lattice following the degradation of the core by light/electron beam irradiation. As a note, it is well known that silver halides are photosensitive materials that are quickly degraded under light/electron beam irradiation.[23-25] Because of this degradation, the shape of the Ag-rich domain observed under the microscope might deviate significantly from that of the void region.

An indirect proof of the core@shell structure was provided by syntheses that were stopped only 10 seconds after the injections of benzoyl bromide. In those cases, the recovered NCs consisted of pure AgBr NCs, as assessed by XRD (**Figure S4**). Also in these cases it was difficult to assess the pristine shape of the AgBr NCs, as they appeared partially degraded to metallic Ag under TEM (**Figure S4**). Only if the reactions were run for at least 30 seconds, were core@shell NCs recovered (**Figure S5**). The shell thickness could be tuned by adjusting the reaction time, but only within a ~16 - 34 nm range (**Figure S5**).

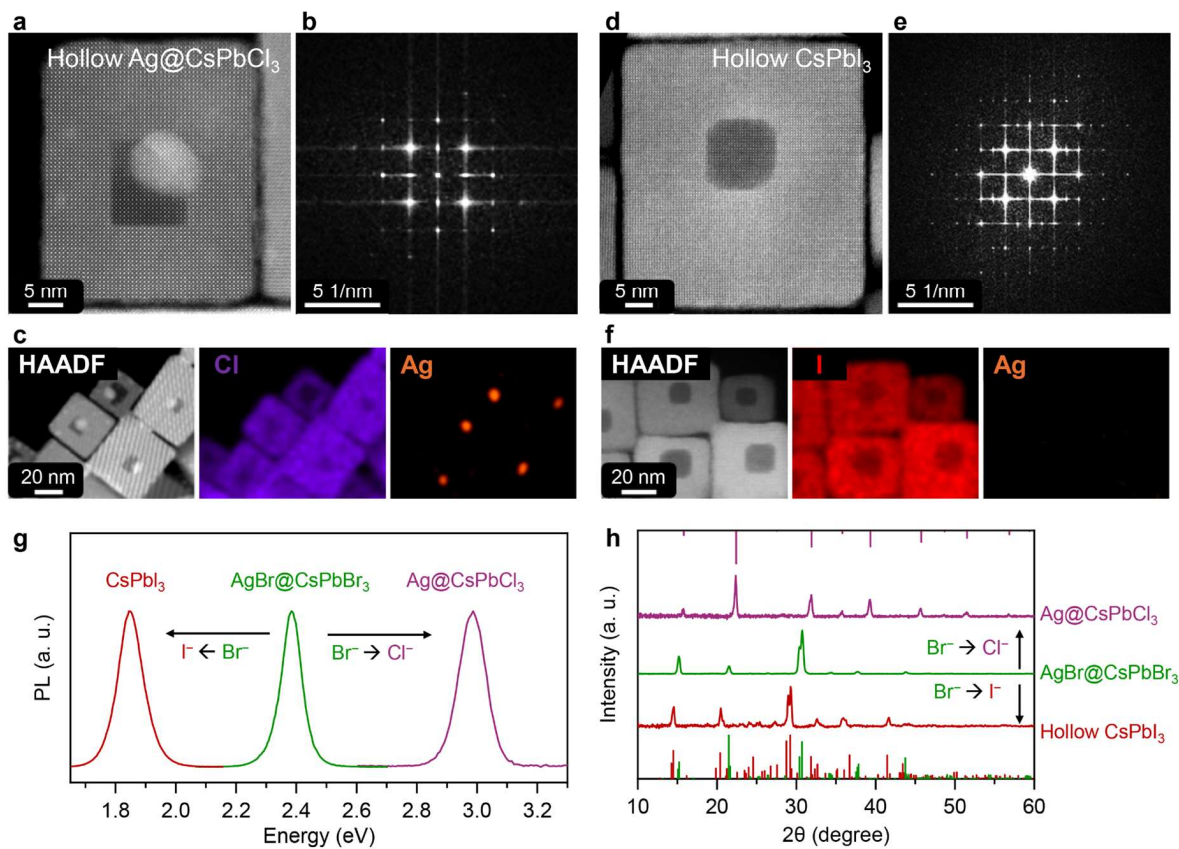

**Figure 2.** Characterization of Ag@CsPbCl$_3$ and CsPbI$_3$ hollow cubes. (a, d) STEM-HAADF images and (b, e) corresponding FFT patterns, and (c, f) STEM-EDX elemental maps of (a-c) Ag@CsPbCl$_3$ and (d-f) CsPbI$_3$ hollow cubes. Unlike for Cl$^-$ exchange (c), no Ag signal is present after I$^-$ exchange (f). (g) PL spectrum and (h) XRD patterns of hollow Ag@CsPbCl$_3$ and CsPbI$_3$ nanocubes (reference marks: red - CsPbI$_3$ (ICSD number 69423), purple - CsPbCl$_3$ (ICSD number 29072)).

Notably, despite variations in overall cube size of the final NCs, the size of the cavity (presumably fully occupied by an AgBr core prior to degradation) was ~12 nm across all the batches, based on (S)TEM analyses. All these experiments and observations agree with a growth process that starts with the nucleation of monodisperse AgBr cores, followed by the overgrowth of a perovskite shell. Also, reactions carried out in the concomitant presence of Ag$^+$ and Zn$^{2+}$ (both in the form of metal oleates) led to a narrower size distribution of the final NCs than those prepared in the sole presence of Ag$^+$ (**Figures S3 and S6**). According to previous works, Zn$^{2+}$ ions are not incorporated in the CsPbBr$_3$ NCs, although they can influence their growth.[7,21] Therefore, all analyses discussed in this work (including those of **Figure 1**) were carried out only on the samples prepared in the concomitant presence of Ag$^+$ and Zn$^{2+}$, unless otherwise stated. We also carried out a series of syntheses aimed at tuning the size of the AgBr core and CsPbBr$_3$ shell, by varying the amounts of Ag$^+$ and Zn$^{2+}$ ions and the reaction temperature. They are discussed in more detail in the experimental section and in the SI (Figures **S7-S9**). These experiments were unsuccessful in achieving a finer control over the geometric parameters in the core@shell NCs.

**Halide exchange reactions.** The core@shell NCs were then subjected to post-synthesis anion exchange with either Cl$^-$ or I$^-$. Previous studies have demonstrated the rapid anion exchange occurring on CsPbX$_3$ (X=Cl, Br, I) NCs (5-20 nm in size).[26,27] In the current work, the Cl$^-$ and I$^-$ exchanges were performed using a lead halide salt (PbCl$_2$ and PbI$_2$) dissolved in a mixture of oleylamine and oleic acid as the halide source (**see details in the Experimental Section**). Cl$^-$ exchange on the NCs delivered hollow CsPbCl$_3$ cubes with an Ag-rich domain that, according to STEM-HAADF and STEM-EDX elemental mapping, was still localized in the cavity (**Figure 2a-c**, see also **Figure S10**). Based on ICP-OES analysis, most of the Ag had been retained inside the particles: the Ag:Pb atomic ratio was 0.07, to be compared to 0.1 of the starting AgBr@CsPbBr$_3$ nanocubes (**Table S2**). The PL spectrum from this sample displayed a characteristic CsPbCl$_3$ emission peak at 2.98 eV (**Figure 2g**), while the XRD pattern confirmed the cubic CsPbCl$_3$ phase, with no indication of metallic Ag peaks, which, again, would be challenging to detect due to the small domain size and volume fraction (**Figure 2h**). Exchange with I$^-$ on the other hand produced hollow CsPbI$_3$ cubes, with no Ag domains in the cavities (**Figure 2d-f and Figure S11**). Experiments carried out to monitor the gradual Br$^-$ to I$^-$ exchange on the AgBr@CsPbBr$_3$ NCs revealed that the core was already partially dissolved when a small amount of I$^-$ was added (I/Br = 0.2) and was completely dissolved at higher I$^-$ loadings (I/Br = 0.4) (**Figure S12**). STEM-HAADF imaging and FFT analysis of the hollow cubes confirmed the orthorhombic CsPbI$_3$ phase (**Figure 2d, e**). STEM-EDX revealed complete I$^-$ exchange, with the signals from Cs, Pb, and I uniformly distributed in the NCs, and no presence of Ag (**Figure 2f**). ICP-OES analysis of the supernatant after the I⁻ exchange reaction confirmed that all of Ag⁺ ions originally located in the NCs had been released into the solution (**Table S3**). The PL spectrum of these hollow NCs (**Figure 2g**) had an emission peak at 1.85 eV, consistent with the $CsPbI_3$ phase, as also corroborated by XRD (**Figure 2h**). The sample had a PLQY of 46%. Notably, in these NCs the hollow region had a truncated cubic shape, different from the $Ag@CsPbCl_3$ hollow cubes discussed above in which the hollow region was cubic (**Figure 2a, d**).

From our experiments we conclude that the AgBr (or Ag) cores in the starting core@shell NCs were completely dissolved during the I⁻ exchange reaction but were unaffected by the Cl⁻ exchange. To validate these findings, we synthesized AgBr and Ag NCs and exposed them to either Cl⁻ or I⁻ ions, under the same reaction conditions of the halide exchange reactions. Adding Cl⁻ to a solution of AgBr NCs had no effect on them, as the NCs preserved their starting AgBr phase (**Figure S13a-c**). Similarly, adding Cl⁻ to a solution of Ag NCs had no major effect other than inducing their aggregation (**Figure S13d-f**). Hence, the Cl⁻ ions neither facilitate Br⁻ to Cl⁻ exchange in AgBr NCs nor they dissolve the Ag NCs. Thus, the only side reaction that could occur in the starting core@shell NCs when treated with Cl⁻ ions was the (further) reduction of photosensitive AgBr core region to metallic Ag during sample handling under light. The same experiments performed by adding I⁻ ions to either AgBr or Ag NCs verified that such addition completely dissolved them (**Figure S14**).

Previous studies have shown that I⁻ and Ag⁺ form stable silver iodide complexes.[28-30] Also, I⁻ ions and molecular iodine can dissolve metallic Au,[31-33] and few studies have also reported the ability of I⁻ ions to dissolve metallic Ag.[34-36] This was rationalized by the chemisorption of I⁻ ions on the surface of the metal particles, which raises their Fermi level and promotes the electron transfer from the metal particles to scavenger species (such as $O_2$) and the concomitant release of Au/Ag metal ions in solution. In our case, it is evident that, as soon as the I⁻ ions diffusing in the NCs reach the Ag/AgBr core, they trigger its dissolution. For metallic Ag, we hypothesize that this occurs most likely through adsorption of I⁻ ions to the surface of the Ag particles, promoting transfer of electrons from Ag to the surrounding environment (the perovskite lattice in this case). These electrons will then find their way out of the NCs and get scavenged. The release of electrons causes the formation of Ag⁺ ions, which can easily diffuse through the perovskite lattice and from there they can reach out to the solution phase. The halide perovskite lattice is indeed known to be "permeable" to various ionic species. Ag⁺ ions, in particular, are capable of diffusing in halide perovskites through interstitial sites.[37-40] Finally, another aspect to consider is that AgBr and AgCl have the same crystal structures and are compatible with the $CsPbBr_3$ (or $CsPbCl_3$) lattice, while AgI has hexagonal crystal structure and is therefore not compatible with the $CsPbI_3$ perovskite lattice.[41, 42] Hence, a hypothetical $AgI@CsPbI_3$ core@shell structure would be unstable in any case.

The hollow $CsPbI_3$ NCs prepared in the previous step could then be used as starting material to prepare hollow $CsPbBr_3$ and $CsPbCl_3$ NCs by halide exchange (**Figure 3**). These reactions were done sequentially. First, a complete exchange with Br⁻ yielded the sample for which STEM-HAADF images are reported in **Figure 3b-d** and PL and XRD in **Figure 3h,i** (green traces). On this sample, a quantitative exchange with Cl⁻ led to the sample which features are reported in **Figure 3e-g** (STEM-HAADF) and in **Figure 3h,i** (PL and XRD, purple traces). Additional STEM data for the two exchange reactions are presented in **Figures S15-S16**. We remark that, while the cavity in the $CsPbI_3$ NCs preserved its truncated cubic shape after Br⁻ exchange, it transformed back to a cubic shape after Cl⁻ exchange, an aspect that is supportive of a perovskite lattice being able to get partly reorganized and that will require further investigation.

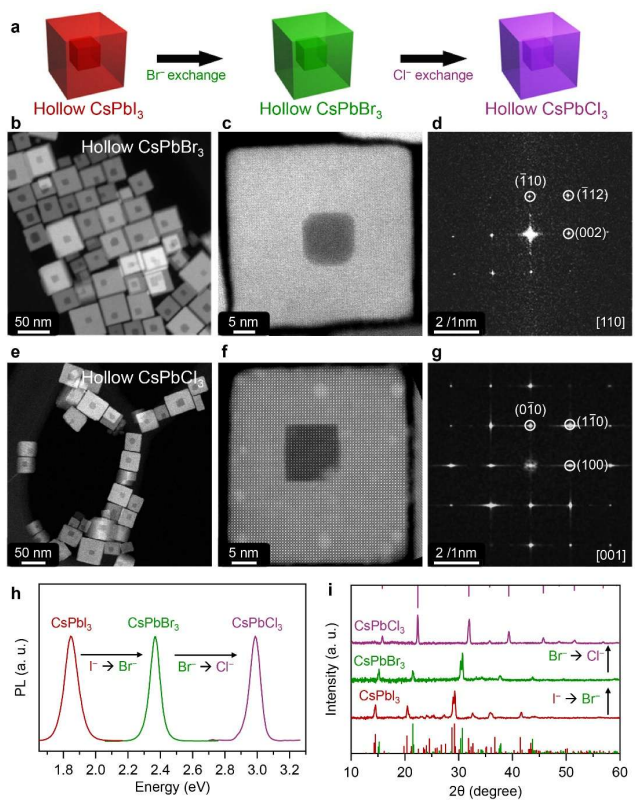

**Figure 3**. Characterization of hollow $CsPbBr_3$ and $CsPbCl_3$ NC. (a) Sketch of Br⁻ and Cl⁻ exchange on $CsPbI_3$ NCs. (b, c, e, f) STEM-HAADF images and (d, g) corresponding FFT images of (b-d) $CsPbBr_3$ and (e-g) $CsPbCl_3$ hollow NCs. (h) PL spectra and (i) XRD patterns of hollow $CsPbCl_3$, $CsPbBr_3$ and $CsPbI_3$ NCs.

**Photophysics of $CsPbI_3$, $CsPbBr_3$ and $CsPbCl_3$ Hollow Cubes.** The PL spectra of $CsPbI_3$, $CsPbBr_3$ and $CsPbCl_3$ hollow NCs (**Figure 3h**) displayed narrow PL peaks (FWHM~74-86 meV) at 1.85 eV, 2.37 eV and 2.99 eV consistent with weakly confined particles.[43-45] The hollow $CsPbBr_3$ cubes, due to their large size and poor dispersibility in solution, exhibited a PL quantum yield (PLQY) of approximately 25%. However, this value is significantly higher than that of the core@shell $AgBr@CsPbBr_3$ cubes, which had a PLQY of only 2.5%, suggesting that the initial presence of AgBr or Ag in the cubes substantially quenches the PL from $CsPbBr_3$. The hollow $CsPbCl_3$ cubes on the other hand had a very low PLQY (< 1%). Post-treatment of conventional $CsPbBr_3$ NCs with specific ligands, such as DDAB, can significantly enhance their PLQY.[46, 47] Following this approach, we treated the hollow CsPbBr$_3$ NCs with DDAB, resulting in PLQY values increase from 25% to 60%.

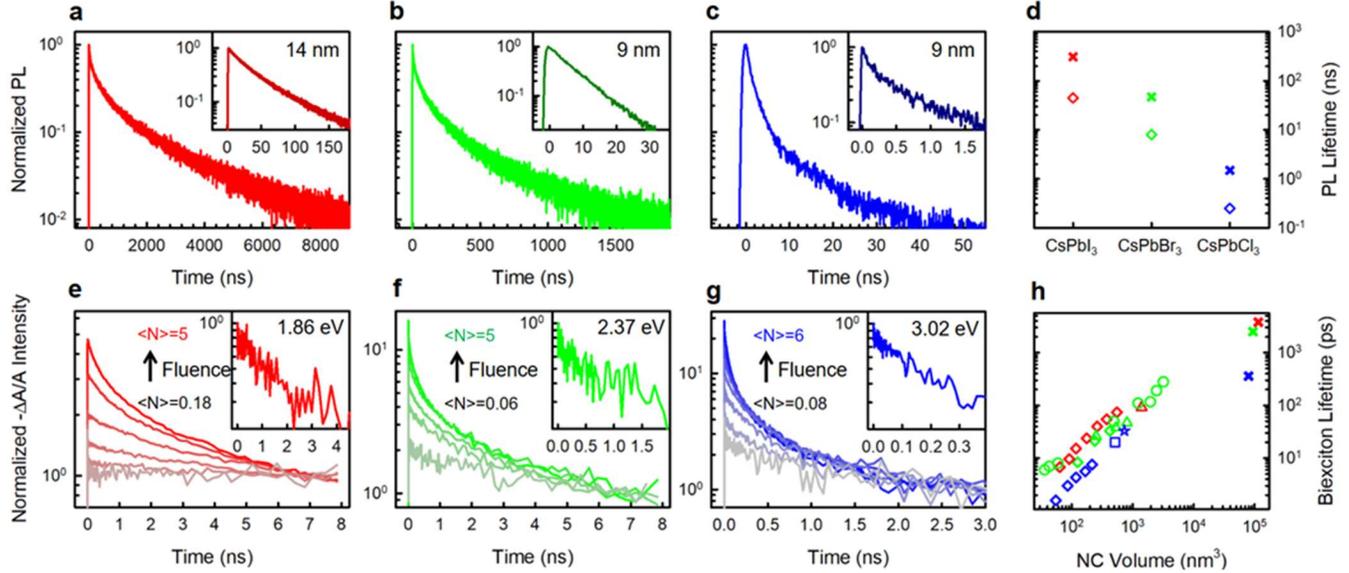

**Figure 4.** (a, b, c) PL dynamics of CsPbX$_3$ (X = I, Br, Cl from left to right) hollow NCs at low excitation fluence. Inset: PL dynamics of medium-confinement, non-hollow CsPbX$_3$ (X = I, Br, Cl from left to right) NCs synthesized with standard routes. (d) PL lifetime for medium-confined (diamonds) non-hollow NCs and for hollow NCs (crosses). (e, f, g) Transient absorption dynamics for CsPbX$_3$ (X = I, Br, Cl from left to right) hollow NCs. Fluence is increasing from grey lines (⟨N⟩=0.1 exc/NC) to coloured lines (⟨N⟩~5 exc/NC). Inset: Extracted biexciton component. (h) Biexciton lifetimes for non-hollow CsPbX$_3$ NCs of different sizes, adapted from Ref. [48] (triangles), [49] (circles), [50] (diamonds), [51] (square), [52] (star) and for the hollow CsPbX$_3$ NCs of this study (crosses).

The photophysics of the CsPbX$_3$ (X=Cl, Br, I) hollow NCs was first investigated at vanishingly low excitation fluence (<100 nJ/cm$^2$) to ensure a single exciton photophysics. PL decay dynamics (**Figure 4a-c**) was multi-exponential, with an initial fast portion followed by a long-lived tail commonly ascribed to regenerated excitons from shallow traps.[53] The effective lifetime (extracted as the time at which the intensity dropped by a factor $e$) of the initial portion was $\tau_{eff}$=310 ns, 47 ns and 1.5 ns for CsPbI$_3$, CsPbBr$_3$ and CsPbCl$_3$ hollow NCs, respectively. These decay times are longer than analogous medium-confinement, non-hollow NCs synthesized with standard routes (insets of **Figure 4a-c and 4d**), an observation that appears to be consistent with enhanced s-p hybridization in large cubes, which partially prohibits the optical transition.[54] Beyond the single exciton photophysics, the particle size also influences the multiexciton dynamics, owing to volume scaling of Auger recombination.[55] To investigate this, we performed transient absorption (TA) measurements at increasing excitation fluences. The TA dynamics of the 1S bleach of the CsPbX$_3$ (X = Cl, Br, I) hollow NCs normalized to their long-time tails are shown in **Figure 4e-g** at increasing excitation fluence spanning average exciton occupancy (0.1≤N≤5). In all cases, increasing the fluence led to the gradual intensification of the initial fast component indicative of multi-excitons. By subtracting the $N \approx 0.1$ to the $N \approx 1$ trace, we extracted the biexciton (XX) lifetime[48] and obtained values of $\tau_{XX}$=3.8 ns (CsPbI$_3$), 2.5 ns (CsPbBr$_3$) and 0.36 ns (CsPbCl$_3$), see **Figure 4h**. These results are in line with the previously reported XX lifetimes for large NCs.[48-52] In all cases, the measured $\tau_{XX}$ values are consistent with the corresponding trend with particle volume.[55] Overall, the PL decay dynamics and TA measurements suggest that the photophysics of the hollow NCs is largely determined by their size, with no apparent effect of the inner hollow core.

CONCLUSIONS

In summary, we have developed a synthesis of AgBr@CsPbBr$_3$ nanocubes by a one-pot approach that exploits the fast nucleation of AgBr seeds followed by the epitaxial growth of a CsPbBr$_3$ shell. Their subsequent reaction with I$^-$ ions completely dissolved the AgBr/Ag core while transforming the thick CsPbBr$_3$ shell into CsPbI$_3$, thus providing access to a new mechanism to prepare hollow metal halide nanostructures. Then, by sequential exchange with Br$^-$ and Cl$^-$ ions, the corresponding hollow CsPbBr$_3$ and CsPbCl$_3$ nanocubes could be prepared. Photoluminescence decay dynamics and transient absorption measurements of the hollow CsPbX$_3$ (X=Cl, Br, I) nanocubes indicate that their photophysics is primarily determined by size, with no discernible effect from the inner cavity, likely due to the small volume fraction of the cavity in these large particles. This work expands the morphological diversity of perovskite nanocrystals that is accessible by colloidal synthesis routes and provides a simple and effective pathway for designing hollow perovskite nanocrystals for potential optoelectronic applications.

EXPERIMENTAL SECTION

**Chemicals.** Hexadecane (99%), oleic acid (OA, technical grade, 90%), cesium(I) carbonate ($Cs_2CO_3$, 98%), lead (II) acetate trihydrate ($Pb(OAc)_2 \cdot 3H_2O$, 99.99%), silver acetate (AgOAc, 99.9%), silver nitrate ($AgNO_3$, 99.9%), zinc (II) acetate ($Zn(OAc)_2$, 99%), sodium citrate dihydrate ($Na_3C_6H_5O_7 \cdot 2H_2O$, >99%), Sodium borohydride ($NaBH_4$, 99%), toluene (anhydrous, 99.8%), oleylamine (OLAm, 98%), 1-dodecanethiol (98%), ethanol (99.8%), methanol (99.8%), didodecyldimethylammonium bromide (DDAB, 98%) were purchased from Sigma-Aldrich. Benzoyl bromide (98%), didodecylmethyl amine (DDMA, >85%) were purchased from Tokyo Chemical Industry (TCI). All reagents were used as received without any further experimental purification.

**Synthesis of core@shell AgBr@CsPbBr$_3$ NCs.**

*Synthesis of core@shell AgBr@CsPbBr$_3$ cubes:* 0.05 mmol of $Cs_2(CO_3)$, 0.1 mmol of $Pb(OAc)_2 \cdot 3H_2O$, 0.2 mmol of Ag(OAc), and 0.025 mmol of $Zn(OAc)_2$ were dissolved in a mixture of 1.5 ml (4.73 mmol) of oleic acid and 6 ml of hexadecane in a flask. The resulting mixture was pumped to vacuum at room temperature for 30 min and at 100 °C for 50 min. The mixture was subsequently placed under a nitrogen atmosphere, and the temperature was raised to 125 °C for 10 minutes to achieve a transparent solution. The solution was then cooled down to 100 °C and a benzoyl bromide solution (obtained by mixing 50 μL of benzoyl bromide (0.42 mmol) in 500 μL of hexadecane) was swiftly injected, triggering nucleation and growth of the NCs. After 1 min of reaction, a tertiary amine solution (obtained by mixing 100 μL DDMA (0.2 mmol) dispersed in 900 μL hexadecane) was swiftly injected, and the reaction was quenched within 10 sec by rapidly cooling it down to room temperature using an ice-water bath. The crude solution was precipitated upon centrifugation at 6000 rpm for 10 min. The precipitate was redispersed in 4 ml of anhydrous toluene. Then the resulting dispersion was centrifuged at 2000 rpm for 5 min. The supernatant was discarded, and the precipitate was redispersed in 4 ml of anhydrous toluene. The final solution was stored in a nitrogen filled glovebox under dark conditions for further characterizations.

*Attempts to optimize reaction conditions*: different amounts of $Zn(OAc)_2$ (0.025, 0.05 and 0.1 mmol), different amounts of Ag(OAc) (0.15, 0.2, 0.25, and 0.30 mmol), and different reaction temperature (90 °C, 100 °C and 110 °C) were tested, following the same reaction protocol of core@shell AgBr@CsPbBr$_3$ cubes discussed above. Increasing the $Zn^{2+}$ amount to 0.05 mmol delivered larger nanocubes. Larger $Zn^{2+}$ amounts (0.1 mmol) resulted in even larger particles, along with AgBr NCs (**Figure S7**). Amounts of $Ag^+$ ions lower than the optimal value of 0.2 mmol (0.15 mmol) led to a mixture of AgBr@CsPbBr$_3$ NCs and CsPbBr$_3$ "only" nanocubes. Amount of $Ag^+$ ions over the optimal value of 0.2 mmol (0.25 mmol) led to broad size distributions and aggregation effects (**Figure S8a, b**). Even larger $Ag^+$ amounts led to mixtures of AgBr and other products ($PbBr_2$ and $CsNO_3$), with no CsPbBr$_3$ (and no PL, **Figure S8c**). Working at lower reaction temperature (90 °C instead of 100 °C), to slow down the growth rate of the NCs and hopefully get a thinner CsPbBr$_3$ shell, were equally unsuccessful and lead again to broad size distributions (**Figure S9**). Higher temperature (110 °C) led instead to inhomogeneous samples composed of a mixture of large (~100 nm) cubes and much smaller NCs.

**Synthesis of halide (Cl⁻, Br⁻, I⁻) precursor solutions for halide exchange reactions.** The halide (Cl⁻, Br⁻, I⁻) precursor solutions were prepared by loading 0.5 mmol of lead halide ($PbCl_2$, $PbBr_2$ and $PbI_2$), 2.5 ml of OA, 2.5 ml of OLAm and 15 ml hexadecane into a 40 ml vial and placing it into an aluminum block on top of a hotplate. The mixture was pumped to vacuum at room temperature for 30 min and at 100 °C for 30 min. Then the temperature was increased to 150 °C until the lead halide salt was dissolved (~15-20 minutes). The final mixture was cooled down to room temperature and stored in a nitrogen filled glovebox.

**Synthesis of hollow CsPbX$_3$ (X=Cl, Br, I) cubes.**

*1) Synthesis of hollow CsPbI$_3$ cubes*: 0.5 mL of a AgBr@CsPbBr$_3$ suspension (15 mM in Br⁻) was dispersed in 2 mL of toluene, then 0.5 mL of iodine precursor (50 mM in I⁻) solution was injected under vigorous stirring. After 60 min (complete anion exchange was confirmed by monitoring the photoluminescence, which at the end of the reaction could be related to the pure CsPbI$_3$ phase), the crude solution was precipitated upon centrifugation at 6000 rpm for 10 min. The precipitate was redispersed in 0.5 ml of toluene and stored in a nitrogen filled glovebox under dark conditions for further characterizations.

*2) Synthesis of hollow CsPbBr$_3$ cubes*: 0.5 mL of a hollow CsPbI$_3$ cubes suspension (15 mM in I⁻, the I⁻ concentration was estimated considering that it should be three times that of $Pb^{2+}$, the latter quantified by ICP-OES analysis) was dispersed in 2 mL of toluene, then 0.5 mL of bromide precursors (50 mM in Br⁻) solution was injected under vigorous stirring. After 60 min anion exchange reaction, the crude solution was precipitated upon centrifugation at 6000 rpm for 10 min. The precipitate was redispersed in 0.5 ml of anhydrous toluene and stored in a nitrogen filled glovebox under dark conditions for further characterizations.

*3) Synthesis of hollow CsPbCl$_3$ cubes*: 0.5 mL of a hollow CsPbBr$_3$ cubes suspension (15 mM in Br⁻) was dispersed in 2 mL of toluene, then 0.5 mL of chloride precursors (50 mM in Cl⁻) solution was injected under vigorous stirring. After 90 min anion exchange reaction, the crude solution was precipitated upon centrifugation at 6000 rpm for 10 min. The precipitate was redispersed in 0.5 ml of anhydrous toluene and stored in a nitrogen filled glovebox under dark conditions for further characterizations.

*4) Synthesis of core@shell Ag@CsPbCl$_3$ cubes*: 0.5 mL of a AgBr@CsPbBr$_3$ suspension (15 mM in Br⁻) was dispersed in 2 mL of toluene, then 0.5 mL of chloride precursors (50 M in Cl⁻) solution was injected under vigorous stirring. After 90 min anion exchange reaction, the crude solution was precipitated upon centrifugation at 6000 rpm for 10 min. The precipitate was redispersed in 0.5 ml of anhydrous toluene and stored in a nitrogen filled glovebox under dark conditions for further characterizations.

**Synthesis of AgBr NCs and their reactions with I⁻ and Cl⁻.** AgBr NCs were synthesized following the procedure published by A. E. Saunders et al.[56] A stock solution was prepared by dissolving silver nitrate (13.4 mg, 0.08 mmol) in 10 mL of toluene containing 170 mg DDAB (0.37 mmol). Sonicating the precursor solution for 60 min resulted in the complete dissolution of the silver nitrate. Next, 1 mL of the stock solution (0.008 mmol $AgNO_3$) was diluted with 1 mL of toluene, then 19 μL of 1-dodecanethiol (0.08 mmol) and 1 mL of methanol was added under vigorous stirring. The solution immediately became turbid. After stirring for 30 seconds, 4 mL of acetone was added and the crude solution was centrifuged upon precipitate the NCs. The precipitate was redispersed in 2 ml of anhydrous toluene and stored in a nitrogen filled glovebox in the dark for further characterizations. $I^-$ and $Cl^-$ addition reactions were performed using the same procedure as for the hollow perovskites. 0.5 mL of a AgBr NCs suspension (~0.002 mmol) was diluted in 0.5 mL of toluene. Then, 0.1 mL of $I^-$ or $Cl^-$ precursor solution (50 mM in $I^-$ or $Cl^-$) was injected under vigorous stirring. After a 30 min reaction, the resulting solution was precipitated upon centrifugation at 6000 rpm for 10 min. Finally, the precipitate was redispersed in 0.5 mL of toluene and stored in a nitrogen filled glovebox in the dark for further characterization.

**Synthesis of Ag NCs and their reactions with $I^-$ and $Cl^-$.** Ag NCs were synthesized according to an optimized procedure published by R. C. Doty et al.[57] An aqueous solution of $NaBH_4$ (0.3 mL, 10 mM) was added to a 10 mL solution containing $AgNO_3$ (0.5 mM) and sodium citrate (0.5 mM) under rapid stirring. The reaction mixture was stirred for 90 min, resulting in a dark color solution, and was then left undisturbed overnight. The crude solution was precipitated upon centrifugation at 12000rpm for 10 min. The precipitate was redispersed in 1 mL of a mixed solvent (ethanol/Milli-Q water = 1/1 (v/v)) and stored in a nitrogen-filled glovebox for further characterization. $I^-$ and $Cl^-$ addition reactions were performed using the same procedure as for the hollow perovskites. Specifically, a 0.3 mL of suspension of Ag NCs (~5 mM Ag) was centrifuged upon 12000 rpm for 10 min, and the resulting precipitate was redispersed in 1 mL of a mixed solvent (ethanol/toluene = 1/2 (v/v)). Then, 0.3 mL of $I^-$ or $Cl^-$ precursor solution (50 mM in $I^-$ or $Cl^-$) was injected under vigorous stirring. After 30 min reaction (the initially dark solution turned colorless upon $I^-$ addition and it remained dark with $Cl^-$ addition), the crude solution was precipitated upon centrifugation at 12,000 rpm for 20 min. Finally, the precipitate was redispersed in 0.5 mL of a mixed solvent (ethanol/Milli-Q water = 1/1 (v/v)) for further characterization.

### Characterization

**X-ray Diffraction (XRD).** XRD analysis was performed on a PANanalytical Empyrean X-ray diffractometer, equipped with a 1.8 kW Cu Kα ceramic X-ray tube (λ = 1.5406 Å) and a PIXcel3D 2 × 2 area detector, operating at 45 kV and 40 mA. Cubes solutions were concentrated in the vial through a flow of nitrogen, then they were drop-cast on a zero-diffraction single crystal silicon substrate. The XRD patterns were collected under ambient conditions.

**Transmission Electron Microscopy (TEM) and Scanning TEM (STEM).** Bright-field TEM (BF-TEM) images with a large field of view were acquired on a JEOL JEM-1400Plus microscope with a thermionic gun ($LaB_6$ crystal) with an acceleration voltage of 120 kV. Cubes solutions were diluted ten times in anhydrous toluene and then drop-cast on the copper TEM grids with an ultrathin carbon film. High-resolution scanning transmission electron microscopy (HRSTEM) images were acquired on a probe-corrected Thermo Fisher Spectra 30-300 STEM operated at 300 kV. Images were acquired on a High-Angle annular Dark Field (HAADF) detector with a current of 30 pA. Compositional maps were acquired using Velox, with a probe current of ~150 pA and rapid rastered scanning by Energy-Dispersive X-ray (EDX) on a Dual-X system comprising two detectors, on either side of the sample, for a total acquisition angle of 1.76 Sr. STEM-HAADF tilt series were acquired by tilting a single tilt tomography sample holder from -70° at 65° with a step of 5° to minimize sample damage. The 3D volume was reconstructed using commercial software (Inspect3D) using the SIRT algorithm.

**Optical Measurements.** The PL spectra of NCs were measured on a Varian Cary Eclipse spectrophotometer ($\lambda_{ex}$ = 350 nm). The photoluminescence (PL) quantum yield (QY) was measured using a FLS920 Edinburgh Instruments spectrofluorimeter equipped with an integrating sphere. The NC samples were dispersed in anhydrous toluene with an optical density of 0.12 at 400 nm, which was the excitation wavelength employed for PLQY measurements, to minimize self-absorption. PL measurements at vanishingly low excitation fluence were performed with a Horiba Triax 190 spectrometer, exciting the samples with a laser ($\lambda_{ex}$ = 360 nm, 405 nm) and collecting the emitted light with a CCD. Time-resolved PL were carried out using a femtosecond amplified laser operated at 20 kHz (see description below), tuned at 360 nm or 405 nm. The emitted light was collected with a phototube coupled to a Cornerstone 260 1/4 m VIS-NIR Monochromator (ORIEL) and a time-correlated single-photon counting unit (time resolution ~200 ps). Ultrafast transient absorption spectroscopy measurements were performed on Ultrafast Systems' Helios TA spectrometer. The laser source was a 10 W Ytterbium amplified laser operated at 1.875 kHz producing ~260 fs pulses at 1030 nm and coupled with an independently tunable optical parametric amplifier from the same supplier that produced the excitation pulses at 360 nm, 400 nm or 500 nm. After passing the pump beam through a synchronous chopper phase-locked to the pulse train (0.938 kHz, blocking every other pump pulse), the pump fluence on the sample was modulated using a variable ND filter. The probe beam was a white light supercontinuum.

**Inductively Coupled Plasma ICP Characterization.** The ICP elemental analysis was carried out via inductively coupled plasma optical emission spectroscopy (ICP-OES) with an iCAP 6300 DUO ICP-OES spectrometer. The samples were first dissolved in 1mL of aqua regia ($HCl/HNO_3$ = 3/1 (v/v)) overnight and diluted with 9 mL of Milli-Q water for measurements.

## ASSOCIATED CONTENT

**Supporting Information**. Calculations on epitaxial structure matching between AgBr and $CsPbBr_3$, additional syntheses of nanocrystals prepared for control experiments and related analyses, additional compositional and structural analyses.

## AUTHOR INFORMATION

### Corresponding Author

* Giorgio Divitini (giorgio.divitini@iit.it)
* Sergio Brovelli (sergio.brovelli@unimib.it)
* Liberato Manna (liberato.manna@iit.it)

### Author Contributions

The manuscript was written through contributions of all authors.

### Funding Sources

I.I, S.B. and G.D. acknowledge funding from the Italian Space Agency (Agenzia Spaziale Italiana, ASI) in the framework of the Research Day "Giornate della Ricerca Spaziale" initiative through the contract ASI N. 2023-4-U.0. L.M. acknowledges funding from European Research Council through the ERC Advanced Grant NEHA (grant agreement n. 101095974). S.T. and L.M. acknowledge for funding the Project IEMAP (Italian Energy Materials Acceleration Platform) within the Italian Research Program ENEA-MASE (Ministero dell'Ambiente e della Sicurezza Energetica) 2021–2024 "Mission Innovation" (agreement 21A033302 GU no. 133/5-6-2021). L. Z. acknowledge fellowship from Horizon Europe through the MSCA-PF agreement no. 101150406 (SPUREPER). A.C and I.K acknowledge funding from European Research Council through the ERC Starting Grant Light-DYNAMO (grant agreement n. 850875).

## REFERENCES

(1) Schmidt, L. C.; Pertegas, A.; Gonzalez-Carrero, S.; Malinkiewicz, O.; Agouram, S.; Minguez Espallargas, G.; Bolink, H. J.; Galian, R. E.; Perez-Prieto, J. Nontemplate Synthesis of $CH_3NH_3PbBr_3$ Perovskite Nanoparticles. *J. Am. Chem. Soc.* **2014**, *136* (3), 850–853.
(2) Kovalenko, M. V.; Protesescu, L.; Bodnarchuk, M. I. Properties and Potential Optoelectronic Applications of Lead Halide Perovskite Nanocrystals. *Science* **2017**, *358* (6364), 745–750.
(3) Fu, Y.; Zhu, H.; Chen, J.; Hautzinger, M. P.; Zhu, X. Y.; Jin, S. Metal Halide Perovskite Nanostructures for Optoelectronic Applications and the Study of Physical Properties. *Nat. Rev. Mater.* **2019**, *4* (3), 169–188.
(4) Dey, A.; Ye, J.; De, A.; Debroye, E.; Ha, S. K.; Bladt, E.; Kshirsagar, A. S.; Wang, Z.; Yin, J.; Wang, Y.; Quan, L. N.; Yan, F.; Gao, M.; Li, X.; Shamsi, J.; Debnath, T.; Cao, M.; Scheel, M. A.; Kumar, S.; Steele, J. A.; Gerhard, M.; Chouhan, L.; Xu, K.; Wu, X.-g.; Li, Y.; Zhang, Y.; Dutta, A.; Han, C.; Vincon, I.; Rogach, A. L.; Nag, A.; Samanta, A.; Korgel, B. A.; Shih, C.-J.; Gamelin, D. R.; Son, D. H.; Zeng, H.; Zhong, H.; Sun, H.; Demir, H. V.; Scheblykin, I. G.; Mora-Seró, I.; Stolarczyk, J. K.; Zhang, J. Z.; Feldmann, J.; Hofkens, J.; Luther, J. M.; Pérez-Prieto, J.; Li, L.; Manna, L.; Bodnarchuk, M. I.; Kovalenko, M. V.; Roeffaers, M. B. J.; Pradhan, N.; Mohammed, O. F.; Bakr, O. M.; Yang, P.; Müller-Buschbaum, P.; Kamat, P. V.; Bao, Q.; Zhang, Q.; Krahne, R.; Galian, R. E.; Stranks, S. D.; Bals, S.; Biju, V.; Tisdale, W. A.; Yan, Y.; Hoye, R. L. Z.; Polavarapu, L. State of the Art and Prospects for Halide Perovskite Nanocrystals. *ACS Nano* **2021**, *15* (7), 10775–10981.
(5) Ye, J.; Gaur, D.; Mi, C.; Chen, Z.; Fernández, I. L.; Zhao, H.; Dong, Y.; Polavarapu, L.; Hoye, R. L. Z. Strongly-Confined Colloidal Lead-Halide Perovskite Quantum Dots: from Synthesis to Applications. *Chem. Soc. Rev.* **2024**, *53* (16), 8095–8122.
(6) Almeida, G.; Goldoni, L.; Akkerman, Q.; Dang, Z.; Khan, A. H.; Marras, S.; Moreels, I.; Manna, L. Role of Acid–Base Equilibria in the Size, Shape, and Phase Control of Cesium Lead Bromide Nanocrystals. *ACS Nano* **2018**, *12* (2), 1704–1711.
(7) Dong, Y.; Qiao, T.; Kim, D.; Parobek, D.; Rossi, D.; Son, D. H. Precise Control of Quantum Confinement in Cesium Lead Halide Perovskite Quantum Dots via Thermodynamic Equilibrium. *Nano Lett.* **2018**, *18* (6), 3716–3722.
(8) Zhang, X.; Wang, Y.; Wu, X.; Wang, F.; Ou, Q.; Zhang, S. A Comprehensive Review on Mechanisms and Applications of Rare-Earth Based Perovskite Nanocrystals. *Chin. J. Chem.* **2024**, *42* (9), 1032–1056.
(9) Tang, X.; Yang, J.; Li, S.; Liu, Z.; Hu, Z.; Hao, J.; Du, J.; Leng, Y.; Qin, H.; Lin, X.; Lin, Y.; Tian, Y.; Zhou, M.; Xiong, Q. Quantum Dots: Single Halide Perovskite/Semiconductor Core/Shell Quantum Dots with Ultrastability and Nonblinking Properties. *Adv. Sci.* **2019**, *6* (18), 1970107.
(10) Li, S.; Lin, H.; Chu, C.; Martin, C.; MacSwain, W.; Meulenberg, R. W.; Franck, J. M.; Chakraborty, A.; Zheng, W. Interfacial B-Site Ion Diffusion in All-Inorganic Core/Shell Perovskite Nanocrystals. *ACS Nano* **2023**, *17* (22), 22467–22477.
(11) Zhao, M.; Wang, X.; Yang, X.; Gilroy, K. D.; Qin, D.; Xia, Y. Hollow Metal Nanocrystals with Ultrathin, Porous Walls and Well-Controlled Surface Structures. *Adv. Mater.* **2018**, *30* (48), 1801956.
(12) Wang, W.; Dahl, M.; Yin, Y. Hollow Nanocrystals through the Nanoscale Kirkendall Effect. *Chem. Mater.* **2013**, *25* (8), 1179–1189.
(13) Song, G.; Han, L.; Zou, W.; Xiao, Z.; Huang, X.; Qin, Z.; Zou, R.; Hu, J. A Novel Photothermal Nanocrystals of $Cu_7S_4$ Hollow Structure for Efficient Ablation of Cancer Cells. *Nano-Micro Lett.* **2014**, *6* (2), 169–177.
(14) Tianou, H.; Wang, W.; Yang, X.; Cao, Z.; Kuang, Q.; Wang, Z.; Shan, Z.; Jin, M.; Yin, Y. Inflating Hollow Nanocrystals through A Repeated Kirkendall Cavitation Process. *Nat. Commun.* **2017**, *8* (1), 1261.
(15) Worku, M.; Tian, Y.; Zhou, C.; Lin, H.; Chaaban, M.; Xu, L. j.; He, Q.; Beery, D.; Zhou, Y.; Lin, X.; Su, Y. f.; Xin, Y.; Ma, B. Hollow Metal Halide Perovskite Nanocrystals with Efficient Blue Emissions. *Sci. Adv.* **2020**, *6* (17), eaaz5961.
(16) Chen, Y.; Zhang, X.; Jiang, J.; Chen, G.; Zhou, K.; Zhang, X.; Li, F.; Yuan, C.; Bao, J.; Xu, X. Microfluidic Synthesis of Hollow $CsPbBr_3$ Perovskite Nanocrystals through The Nanoscale Kirkendall Effect. *Nano Res.* **2024**, *17* (9), 8487–8494.
(17) Lai, M.; Obliger, A.; Lu, D.; Kley, C. S.; Bischak, C. G.; Kong, Q.; Lei, T.; Dou, L.; Ginsberg, N. S.; Limmer, D. T. Intrinsic Anion Diffusivity in Lead Halide Perovskites is Facilitated by A Soft Lattice. *Proc. Natl. Acad. Sci.* **2018**, *115* (47), 11929–11934.
(18) Pan, D.; Fu, Y.; Chen, J.; Czech, K. J.; Wright, J. C.; Jin, S. Visualization and Studies of Ion-Diffusion Kinetics in Cesium Lead Bromide Perovskite Nanowires. *Nano Lett.* **2018**, *18* (3), 1807–1813.
(19) Trivelli, A. P. H.; Sheppard, S. E. On the Visible Decomposition of Silver Halide Grains by Light. *J. Phys. Chem.* **2002**, *29* (12), 1568–1582.

(20) James, T.; Kornfeld, G. Reduction of Silver Halides and the Mechanism of Photographic Development. *Chem. Rev.* **1942**, *30* (1), 1−32.
(21) Li, Z.; Goldoni, L.; Wu, Y.; Imran, M.; Ivanov, Y. P.; Divitini, G.; Zito, J.; Panneerselvam, I. R.; Baranov, D.; Infante, I.; De Trizio, L.; Manna, L. Exogenous Metal Cations in the Synthesis of CsPbBr$_3$ Nanocrystals and Their Interplay with Tertiary Amines. *J. Am. Chem. Soc.* **2024**, *146* (30), 20636−20648.
(22) Toso, S.; Dardzinski, D.; Manna, L.; Marom, N. Structure Prediction of Ionic Epitaxial Interfaces with Ogre Demonstrated for Colloidal Heterostructures of Lead Halide Perovskites. *ACS Nano* **2025**, *19* (15), 5326−5341.
(23) Bachman, P. L. 2.6 SILVER HALIDE PHOTOGRAPHY. *Handbook of Optical Holography* **1979**, 89.
(24) Hamilton, J. The Silver Halide Photographic Process. *Adv. Phys.* **1988**, *37* (4), 359−441.
(25) Huang, K.; Li, C.; Zheng, Y.; Wang, L.; Wang, W.; Meng, X. Recent Advances on Silver-Based Photocatalysis: Photocorrosion Inhibition, Visible-Light Responsivity Enhancement, and Charges Separation Acceleration. *Sep. Purif. Technol.* **2022**, *283*, 120194.
(26) Akkerman, Q. A.; D'Innocenzo, V.; Accornero, S.; Scarpellini, A.; Petrozza, A.; Prato, M.; Manna, L. Tuning the Optical Properties of Cesium Lead Halide Perovskite Nanocrystals by Anion Exchange Reactions. *J. Am. Chem. Soc.* **2015**, *137* (32), 10276−10281.
(27) Nedelcu, G.; Protesescu, L.; Yakunin, S.; Bodnarchuk, M. I.; Grotevent, M. J.; Kovalenko, M. V. Fast Anion-Exchange in Highly Luminescent Nanocrystals of Cesium Lead Halide Perovskites (CsPbX$_3$, X = Cl, Br, I). *Nano Lett.* **2015**, *15* (8), 5635−5640.
(28) Gaizer, F.; Johansson, G. Silver Iodide Complexes in DMSO and DMF Solutions. *Acta Chem. Scand. A* **1988**, *42*, 259−268.
(29) Leden, I. Anionic Silver Iodide Complexes in Aqueous Solution. *Acta Chem. Scand.* **1956**, *10*, 812−821.
(30) Berne, E.; Weill, M. A Study of Silver Iodide Complexes in Water Solutions by Self-Diffusion Measurements. *J. Phys. Chem.* **1960**, *64* (2), 258−261.
(31) Green, T. Gold Etching for Microfabrication. *Gold Bull.* **2014**, *47* (3), 205−216.
(32) Zupanc, A.; Heliövaara, E.; Moslova, K.; Eronen, A.; Kemell, M.; Podlipnik, Č.; Jereb, M.; Repo, T. Iodine-Catalysed Dissolution of Elemental Gold in Ethanol. *Angew. Chem., Int. Ed.* **2022**, *61* (14), e202117587.
(33) Nakao, Y.; Sone, K. Reversible Dissolution/Deposition of Gold in Iodine–Iodide–Acetonitrile Systems. *Chem. Commun.* **1996**, (8), 897−898.
(34) Inukai, J.; Tryk, D. A.; Abe, T.; Wakisaka, M.; Uchida, H.; Watanabe, M. Direct STM Elucidation of the Effects of Atomic-Level Structure on Pt (111) Electrodes for Dissolved CO Oxidation. *J. Am. Chem. Soc.* **2013**, *135* (4), 1476−1490.
(35) Henglein, A. Physicochemical Properties of Small Metal Particles in Solution:" Microelectrode" Reactions, Chemisorption, Composite Metal Particles, and the Atom-to-Metal Transition. *J. Phys. Chem.* **1993**, *97* (21), 5457−5471.
(36) Stefancu, A.; Iancu, S. D.; Coman, V.; Leopold, L. F.; Leopold, N. Tuning the Potential of Nanoelectrodes to Maximum: Ag and Au Nanoparticles Dissolution by I$^-$ adsorption via Mg$^{2+}$ Adions. *Rom. Rep. Phys.* **2021**, *73* (2), 502.
(37) Li, J.; Dong, Q.; Li, N.; Wang, L. Direct Evidence of Ion Diffusion for the Silver-Electrode-Induced Thermal Degradation of Inverted Perovskite Solar Cells. *Adv. Energy Mater.* **2017**, *7* (14), 1602922.
(38) Ming, W.; Yang, D.; Li, T.; Zhang, L.; Du, M. H. Formation and Diffusion of Metal Impurities in Perovskite Solar Cell Material CH$_3$NH$_3$PbI$_3$: Implications on Solar Cell Degradation and Choice of Electrode. *Adv. Sci.* **2018**, *5* (2), 1700662.
(39) Zhou, S.; Ma, Y.; Zhou, G.; Xu, X.; Qin, M.; Li, Y.; Hsu, Y. J.; Hu, H.; Li, G.; Zhao, N. Ag-Doped Halide Perovskite Nanocrystals for Tunable Band Structure and Efficient Charge Transport. *ACS Energy Lett.* **2019**, *4* (2), 534−541.
(40) Yang, W.; Zhu, B.; Hou, Y.; Yang, X.; Pang, J. Ag Diffusion Effect on the Crystal Structure, Band Structure, and Optical Property of α-CsPbI$_3$ Perovskite Materials. *Phys. B Condens. Matter* **2020**, *579*, 411917.
(41) Berry, C. R. Structure and Optical Absorption of AgI Microcrystals. *Phys. Rev.* **1967**, *161* (3), 848−851.
(42) Glaus, S.; Calzaferri, G. The Band Structures of the Silver Halides AgF, AgCl, and AgBr: A Comparative Study. *Photochem. Photobiol. Sci.* **2003**, *2* (4), 398−401.
(43) Eperon, G. E.; Paternò, G. M.; Sutton, R. J.; Zampetti, A.; Haghighirad, A. A.; Cacialli, F.; Snaith, H. J. Inorganic Caesium Lead Iodide Perovskite Solar Cells. *J. Mater. Chem. A* **2015**, *3* (39), 19688−19695.
(44) Mannino, G.; Deretzis, I.; Smecca, E.; La Magna, A.; Alberti, A.; Ceratti, D.; Cahen, D. Temperature-Eependent Optical Band Gap in CsPbBr$_3$, MAPbBr$_3$, and FAPbBr$_3$ Single Crystals. *J. Phys. Chem. Lett.* **2020**, *11* (7), 2490−2496.
(45) Sebastian, M.; Peters, J.; Stoumpos, C.; Im, J.; Kostina, S.; Liu, Z.; Kanatzidis, M.; Freeman, A.; Wessels, B. Excitonic Emissions and Above-Band-Gap Luminescence in the Single-Crystal Perovskite Semiconductors CsPbBr$_3$ and CsPbCl$_3$. *Phys. Rev. B* **2015**, *92* (23), 235210.
(46) Imran, M.; Ijaz, P.; Goldoni, L.; Maggioni, D.; Petralanda, U.; Prato, M.; Almeida, G.; Infante, I.; Manna, L. Simultaneous Cationic and Anionic Ligand Exchange for Colloidally Stable CsPbBr$_3$ Nanocrystals. *ACS Energy Lett.* **2019**, *4* (4), 819−824.
(47) Fiuza-Maneiro, N.; Sun, K.; Lopez-Fernandez, I.; Gomez-Grana, S.; Muller-Buschbaum, P.; Polavarapu, L. Ligand Chemistry of Inorganic Lead Halide Perovskite Nanocrystals. *ACS Energy Lett.* **2023**, *8* (2), 1152−1191.
(48) Makarov, N. S.; Guo, S.; Isaienko, O.; Liu, W.; Robel, I.; Klimov, V. I. Spectral and Dynamical Properties of Single Excitons, Biexcitons, and Trions in Cesium–Lead-Halide Perovskite Quantum Dots. *Nano Lett.* **2016**, *16* (4), 2349−2362.
(49) Fratelli, A.; Zaffalon, M. L.; Mazzola, E.; Dirin, D. N.; Cherniukh, I.; Otero-Martínez, C.; Salomoni, M.; Carulli, F.; Rossi, F.; Meinardi, F. Size-Dependent Multiexciton Dynamics Governs Scintillation From Perovskite Quantum Dots. *Adv. Mater.* **2025**, *37* (5), 2413182.
(50) Li, Y.; Luo, X.; Ding, T.; Lu, X.; Wu, K. Size- and Halide-Dependent Auger Recombination in Lead Halide Perovskite Nanocrystals. *Angew. Chem., Int. Ed.* **2020**, *132* (34), 14398−14401.
(51) Ahumada-Lazo, R.; Alanis, J. A.; Parkinson, P.; Binks, D. J.; Hardman, S. J.; Griffiths, J. T.; Wisnivesky Rocca Rivarola, F.; Humphrey, C. J.; Ducati, C.; Davis, N. J. Emission Properties and Ultrafast Carrier Dynamics of CsPbCl$_3$ Perovskite Nanocrystals. *J. Phys. Chem. C* **2019**, *123* (4), 2651−2657.

(52) Erroi, A.; Carulli, F.; Cova, F.; Frank, I.; Zaffalon, M. L.; Llusar, J.; Mecca, S.; Cemmi, A.; Di Sarcina, I.; Rossi, F. Ultrafast Nanocomposite Scintillators Based on Cd-Enhanced $CsPbCl_3$ Nanocrystals in Polymer Matrix. *ACS Energy Lett.* **2024**, *9* (5), 2333–2342.

(53) Rodà, C.; Fasoli, M.; Zaffalon, M. L.; Cova, F.; Pinchetti, V.; Shamsi, J.; Abdelhady, A. L.; Imran, M.; Meinardi, F.; Manna, L. Understanding Thermal and A-Thermal Trapping Processes in Lead Halide Perovskites Towards Effective Radiation Detection Schemes. *Adv. Funct. Mater.* **2021**, *31* (43), 2104879.

(54) Krieg, F.; Sercel, P. C.; Burian, M.; Andrusiv, H.; Bodnarchuk, M. I.; Stöferle, T.; Mahrt, R. F.; Naumenko, D.; Amenitsch, H.; Rainò, G. Monodisperse Long-Chain Sulfobetaine-Capped $CsPbBr_3$ Nanocrystals and Their Superfluorescent Assemblies. *ACS Cent. Sci.* **2020**, *7* (1), 135–144.

(55) Robel, I.; Gresback, R.; Kortshagen, U.; Schaller, R. D.; Klimov, V. I. Universal Size-Dependent Trend in Auger Recombination in Direct-Gap and Indirect-Gap Semiconductor Nanocrystals. *Phys. Rev. Lett.* **2009**, *102* (17), 177404.

(56) Saunders, A. E.; Popov, I.; Banin, U. Synthesis and Characterization of Organic-Soluble Ag/AgBr Dimer Nanocrystals. *Zeitschrift für anorganische und allgemeine Chemie* **2007**, *633* (13-14), 2414–2419.

(57) Doty, R. C.; Tshikhudo, T. R.; Brust, M.; Fernig, D. G. Extremely Stable Water-Soluble Ag Nanoparticles. *Chem. Mater.* **2005**, *17* (18), 4630–4635.

Supporting Information for:

# Core@Shell AgBr@CsPbBr$_3$ Nanocrystals as Precursors to Hollow Lead Halide Perovskite Nanocubes

Zhanzhao Li[1], Yurii P. Ivanov[2], Anna Cabona[1,3], Andrea Fratelli[1,4], Stefano Toso,[1] Saptarshi Chakraborty[4], Giorgio Divitini[2]*, Ilka Kriegel[3], Sergio Brovelli[4]*, Liberato Manna[1]*

[1] Nanochemistry, Istituto Italiano di Tecnologia, Via Morego 30, 16163, Genova, Italy

[2] Electron Spectroscopy and Nanoscopy, Istituto Italiano di Tecnologia, Via Morego 30, 16163, Genova, Italy

[3] Department of Applied Science and Technology, Politecnico di Torino, Corso Duca degli Abruzzi 34, 10129, Turin, Italy

[4] Dipartimento di Scienza dei Materiali, Università degli Studi di Milano Bicocca, via R. Cozzi 55, 20125, Milano, Italy

Contents: Figures S1-S16, Tables S1-S3, additional references

**Epitaxial structure matching between AgBr and CsPbBr$_3$.**

The likelihood of a successful growth of AgBr@CsPbBr$_3$ core-shell nanoparticles was evaluated by simulating the possible AgBr/CsPbBr$_3$ epitaxial relations, using the Ogre library for the prediction of interfaces between ionic materials. Due to the small deviations of the orthorhombic Pnma CsPbBr$_3$ structure from the Pm-3m cubic perovskite prototype, in our simulations, we initially considered using the latter case in CsPbBr$_3$ structure simulation, which allows for a more intuitive interpretation of results.

The key requirements for growing an epitaxial shell are:

- 1) That the two materials can form stable interfaces along multiple directions of their lattice, to ensure that the seed material can be fully enveloped.
- 2) That these interfaces form by retaining the same relative orientations of the two atomic lattices, to ensure that the outer shell can grow as a single-crystalline domain.

The lattice matching table in **Figure S1** (see Ref.[1] for details on the interpretation) clearly indicates that these conditions can be met for AgBr/CsPbBr$_3$, as both materials are cubic and all interfaces where (hkl)$_{AgBr}$ = (hkl)$_{CsPbBr3}$ display favorable geometric matching with small interface supercells and low strain of 1.16%.

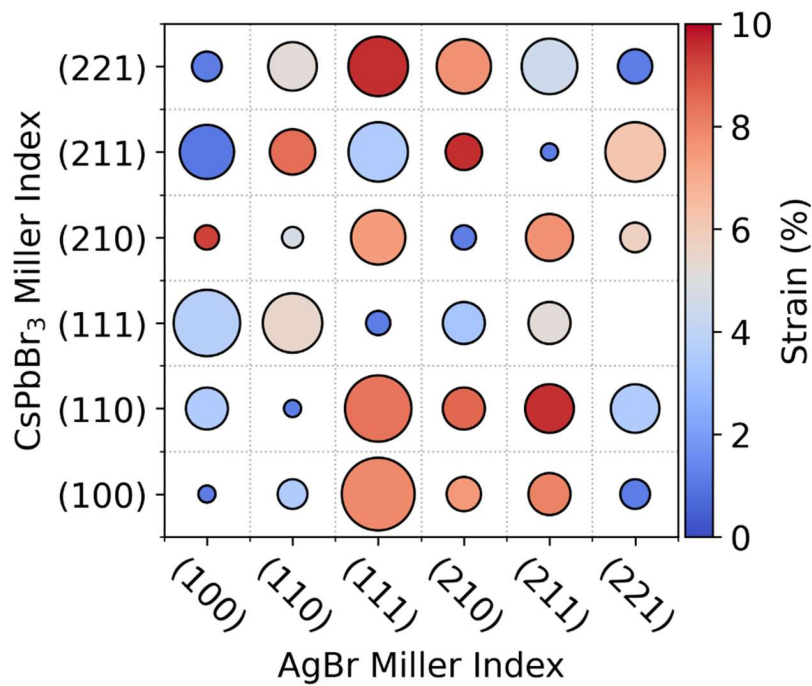

**Figure S1. Lattice matching table for the AgBr/cubic-CsPbBr$_3$ pair.** Each circle summarizes the geometric properties of the corresponding (hkl)/(h'k'l') interface. The circle area indicates the interface supercell surface (smaller = better), while the circle color indicates the interface strain (blue = better).

Simulations of the possible interface structures, performed by considering (hkl) = (h'k'l') up to h,k,l = 2, indicate that all crystallographic directions can give rise to stable interfaces between AgBr and CsPbBr$_3$ (**Table S1**), attesting to their excellent structural compatibility.

**Table S1.** Predicted geometric parameters and interface energies for AgBr/CsPbBr$_3$ interfaces of the (hkl) = (h'k'l') type.

| (hkl) AgBr | (hkl) CsPbBr$_3$ | Strain (%) | Area /Å$^2$ | Interface Energy meV/Å$^2$ |
|---|---|---|---|---|
| (100) | (100) | 1.16 | 33.7 | 5.1 |
| (110) | (110) | 1.16 | 47.7 | 42.0 |
| (111) | (111) | 1.16 | 58.4 | 47.6 |
| (210) | (210) | 1.16 | 150.7 | 22.1 |
| (211) | (211) | 1.16 | 82.6 | 49.8 |
| (221) | (221) | 1.16 | 199.9 | 17.7 |

However, the interface energy for the (100)/(100) AgBr/CsPbBr$_3$ interface (**Figure S2**) is significantly lower than that of any other direction. Interpreting this result in the light of the Wulff construction for crystals with minimal surface energy leads to the conclusion that the most likely shape for an AgBr seed embedded inside a CsPbBr$_3$ domain is that of a sharp cube exposing preeminently its (100) facets.

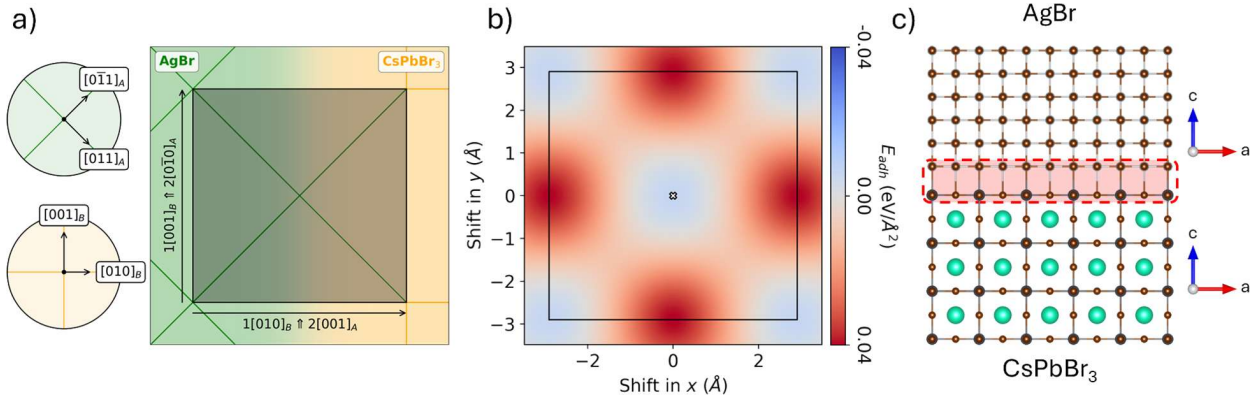

**Figure S2. (100)/(100) AgBr/CsPbBr$_3$ interface.** a) Supercell scheme, showing the relative orientation of the AgBr and CsPbBr$_3$ lattice vectors at the interface. b) Potential energy surface describing the relative in-plane shift of the two materials which leads to the most stable interface configuration. c) Atomic model of the interface as produced by Ogre, showing that the most favorable termination for CsPbBr$_3$ at the interface is the PbBr$_2$ plane, which leads to the formation of Pb-Br and Ag-Br bonds at the interface (highlighted in red).

Control simulations performed considering the orthorhombic Pnma structure for CsPbBr$_3$ instead of the simplified Pm-3m confirmed the same results, with minimal stability differences between the (100)/(010) AgBr/Pnma-CsPbBr$_3$ interface (9.5 meV/Å$^2$) and the (100)/(101) AgBr/Pnma-CsPbBr$_3$ interface (9.3 meV/Å$^2$). In the interest of structural accuracy, an interface model displaying the orthorhombic structure for CsPbBr$_3$ is shown as a part of **Scheme 1b** in the Main text.

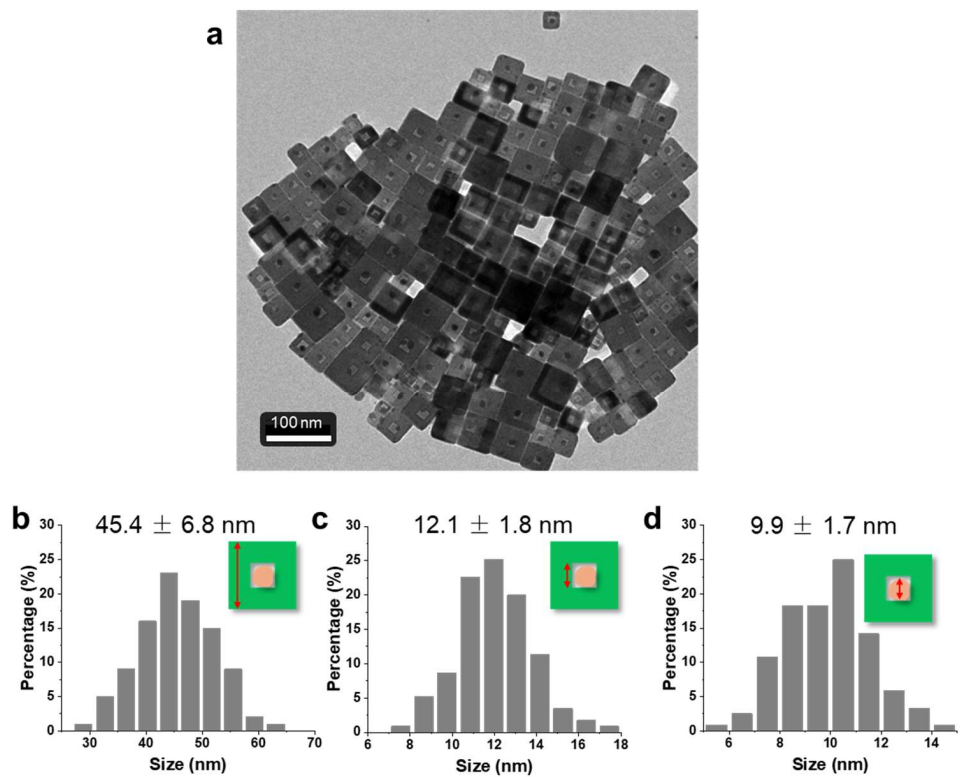

**Figure S3.** (a) TEM image of NCs synthesized with $Zn^{2+}$ addition (molar ratios of Cs:Pb:Ag:Zn = 1:1:2:0.25). (b, c, d) Size distributions of the nanocube, cavity, and solid core region.

**Table S2.** Ag:Pb atomic ratio in AgBr@CsPbBr$_3$, Ag@CsPbCl$_3$ and hollow CsPbI$_3$ NCs according to ICP-OES analysis.

|  | Element content (ppm) | | |
| --- | --- | --- | --- |
| Samples | Pb | Ag | Ag:Pb |
| AgBr@CsPbBr$_3$ | 1.09 | 0.11 | 0.10 |
| Ag@CsPbCl$_3$ | 1.69 | 0.11 | 0.07 |
| Hollow CsPbI$_3$ | 1.4 | 0 | - |

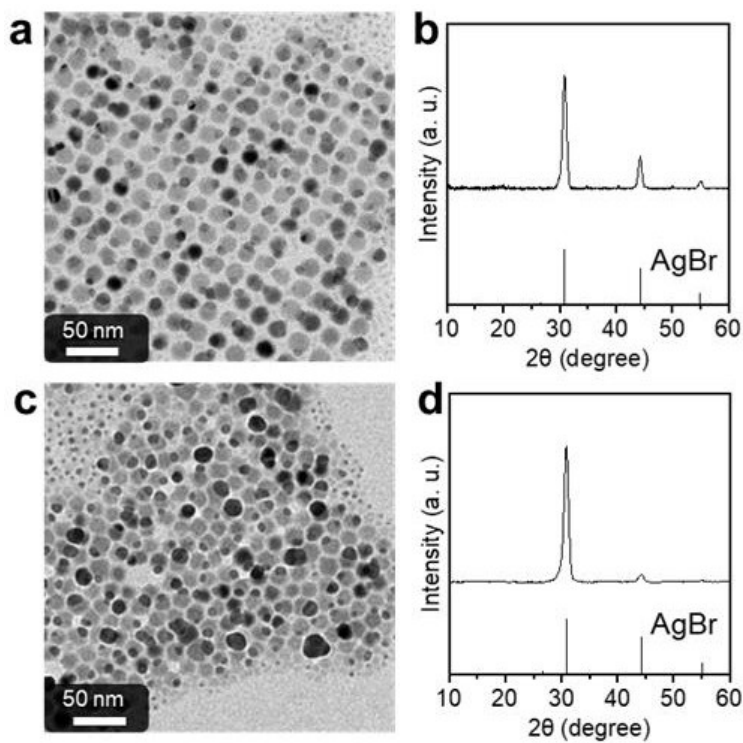

**Figure S4.** (a, c) TEM images, and (b, d) XRD patterns of NCs synthesized (a, b) without $Zn^{2+}$ addition and (c, d) with $Zn^{2+}$ addition (Cs:Pb:Ag:Zn=1:1:2:0.25). In both cases, the synthesis was stopped 10 sec after the injection. In (b) and (c), the tabulated reflections for AgBr (ICSD number 56546) are marked by vertical black bars. In both cases, AgBr NCs were synthesized.

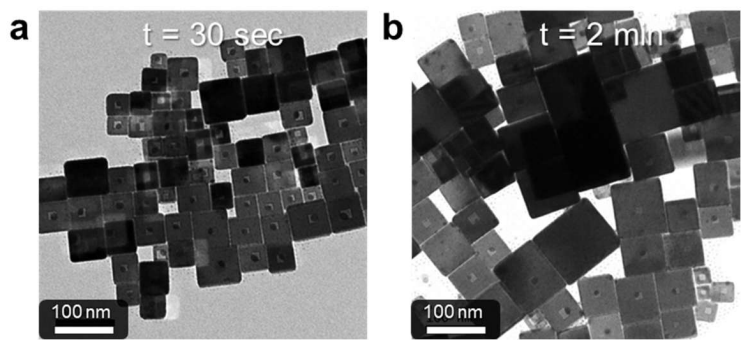

**Figure S5.** TEM images of samples recovered at (a) 30 sec and (b) 2 min after the injection (molar ratios of Cs:Pb:Ag:Zn = 1:1:2:0.25).

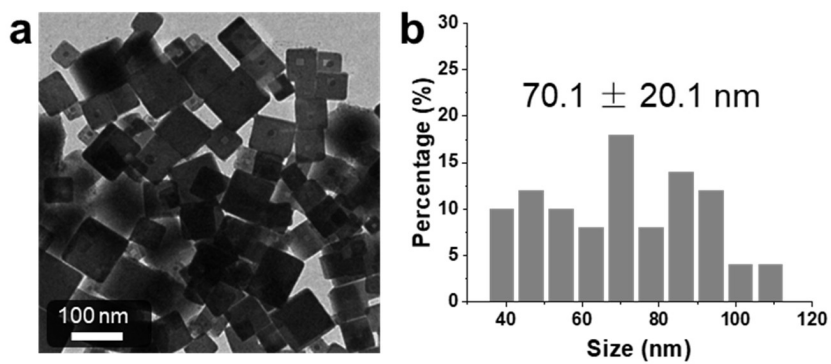

**Figure S6.** (a) TEM images and (b) size distribution of NCs synthesized without $Zn^{2+}$ addition (molar ratios of Cs:Pb:Ag = 1:1:2).

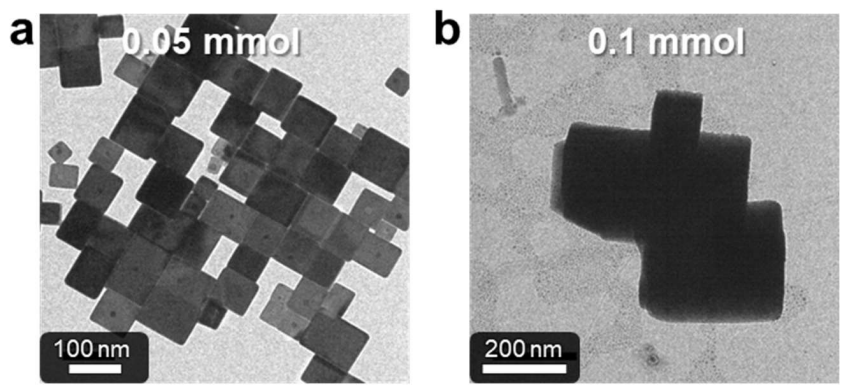

**Figure S7.** TEM images of samples obtained with (a) 0.05 mmol and (b) 0.1 mmol $Zn^{2+}$ addition (molar ratios of Cs:Pb:Ag = 1:1:2).

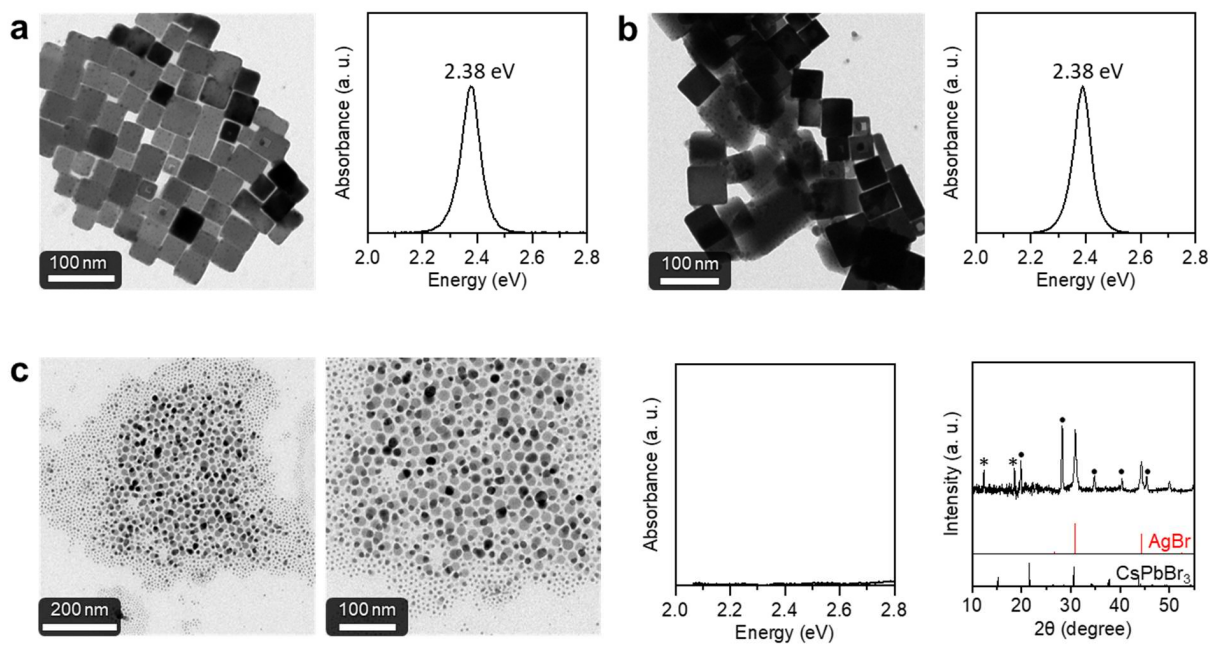

**Figure S8.** TEM, PL spectra and XRD patterns of samples obtained with (a) 0.15 mmol, (b) 0.25 mmol and (c) 0.3 mmol Ag$^+$ addition (molar ratios of Cs:Pb:Zn = 1:1:0.25). In the XRD pattern analysis in panel c, except for the peaks assigned to AgBr, additional peaks are assigned to PbBr$_2$ (marked by *) and AgNO$_3$ (marked by •).

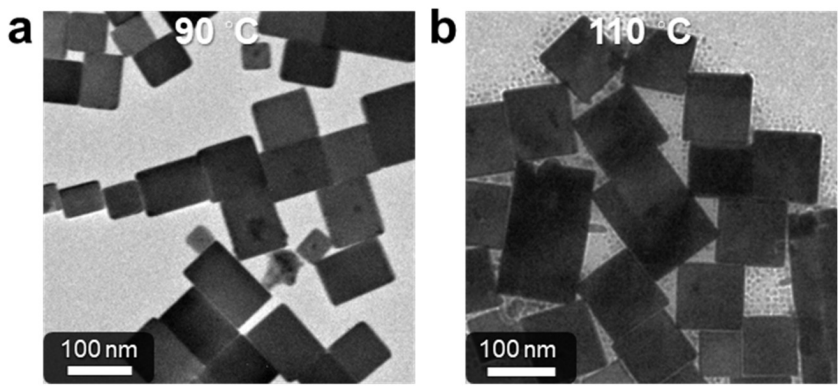

**Figure S9.** TEM images of samples synthesized at reaction temperatures of (a) 90°C and (b) 110°C (molar ratios of Cs:Pb:Ag:Zn = 1:1:2:0.25).

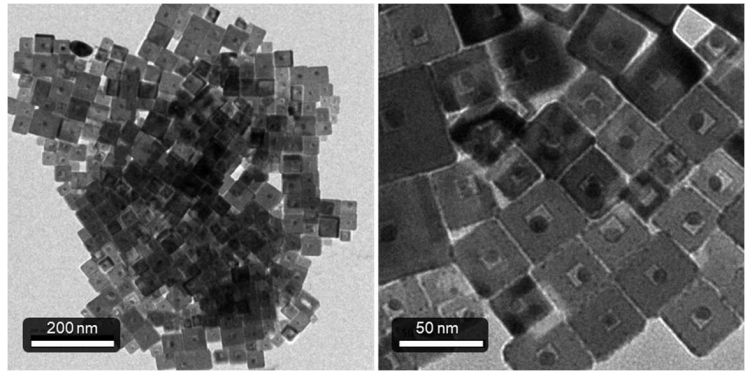

**Figure S10.** TEM images of Ag@CsPbCl$_3$ NCs prepared by Cl$^-$ exchange on the AgBr@CsPbBr$_3$ NCs.

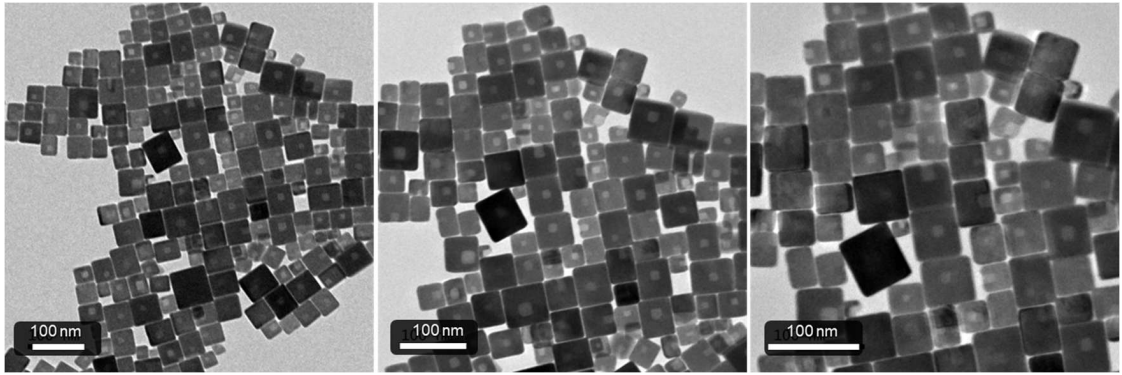

**Figure S11.** TEM images of hollow CsPbI$_3$ NCs prepared by I$^-$ exchange on the AgBr@CsPbBr$_3$ NCs.

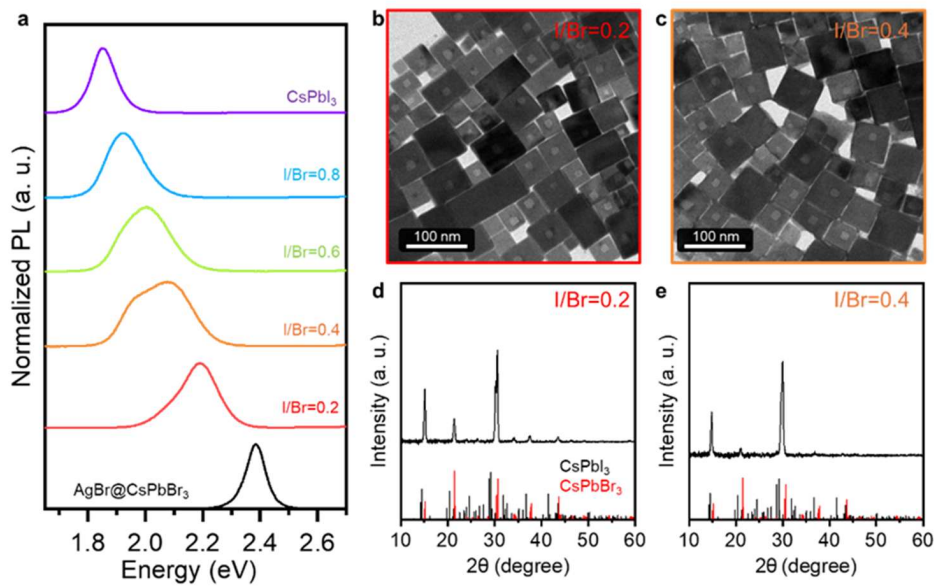

**Figure S12.** Br$^-$ to I$^-$ exchange on core@shell Ag@CsPbBr$_3$ NCs. (a) PL spectra of the starting, intermediate and final samples for the Br$^-$ to I$^-$ exchange reaction. (b, c) TEM images and (d, e) XRD patterns of a couple of intermediate exchange samples: (b, d) I/Br = 0.2 and (c, e) I/Br = 0.4, which resulted from sequential aliquots of iodide being added in the same reaction. In (d, e), the reflections of CsPbBr$_3$ (ICSD number 97851) and CsPbI$_3$ (ICSD number 69423) are represented by vertical black and red bars, respectively.

**Table S3.** ICP-OES analysis of the supernatant solution after the I⁻ exchange reaction on the AgBr@CsPbBr$_3$ NCs. In the I⁻ exchange reaction, 0.5 mL of a colloidal suspension of AgBr@CsPbBr$_3$ NCs (5 mM in Pb) was added to 2 mL of toluene, then 0.5 mL of iodine precursor (25 mM Pb) solution was added. The NCs were then separated and the supernatant was analyzed. The experimental Ag:Pb ratio was 0.018 (0.02/1.09), which is very close to the estimated Ag:Pb ratio of 0.02 in case all the Ag initially present in the AgBr@CsPbBr$_3$ NC was released into the solution and thus in the supernatant.

| Element | Content (ppm) |
|---------|---------------|
| Pb      | 1.09          |
| Ag      | 0.02          |

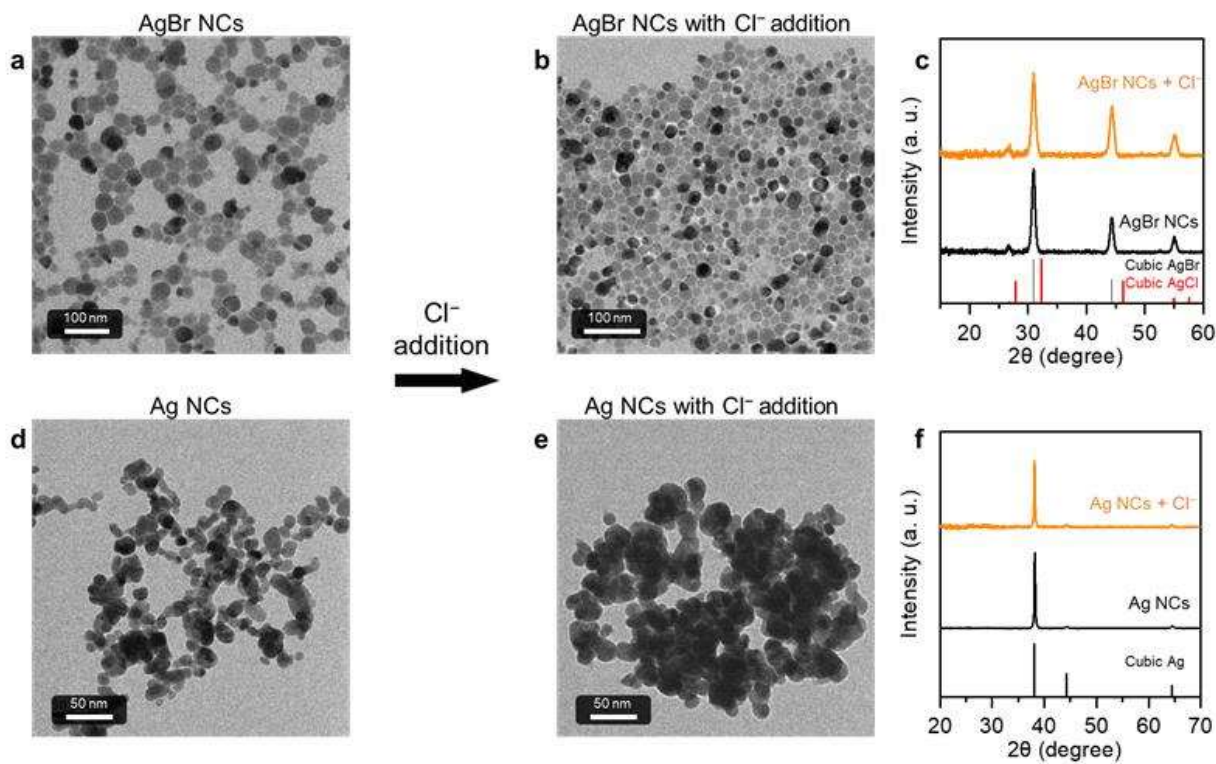

**Figure S13.** Results of Cl⁻ addition to a colloidal suspension of AgBr NCs (top panels) and Ag NCs (bottom panels). (a, b) TEM images and (c) XRD pattens of the (a) starting AgBr NCs and (b) resulting NCs after Cl⁻ addition. (d, e) TEM images and (f) XRD pattens of the (e) starting Ag NCs and (f) resulting NCs after Cl⁻ addition. In (c), the reflections of cubic AgCl (ICSD number 56548) are represented by vertical red bars. In (f), the reflections of Ag (ICSD number 13762) are represented by vertical black bars.

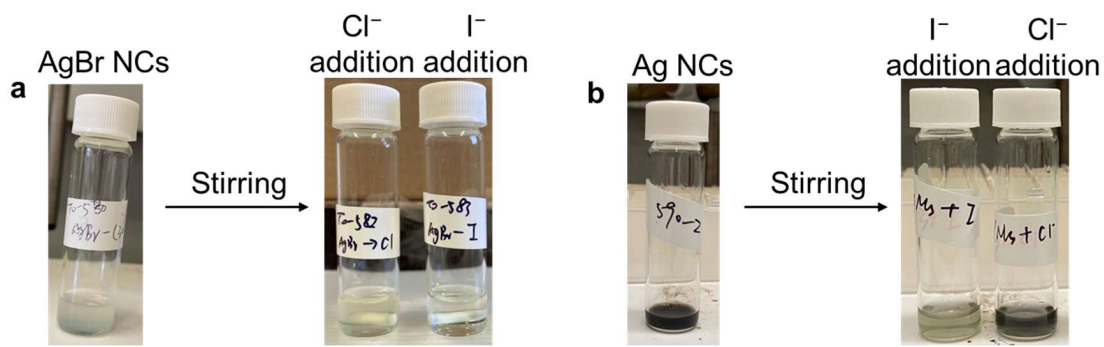

**Figure S14.** (a, b) Photographs of vials before and after adding I⁻ and Cl⁻ to colloidal suspension of AgBr NCs (a) and Ag NCs (b). In the Cl⁻ cases, we identified the AgBr or Ag NCs in the solutions, as discussed in Figure S10. In the I⁻ cases, both suspensions became transparent after I⁻ addition, and no precipitate was recovered after centrifugation. No NCs could be identified by TEM observations of these two latter solutions.

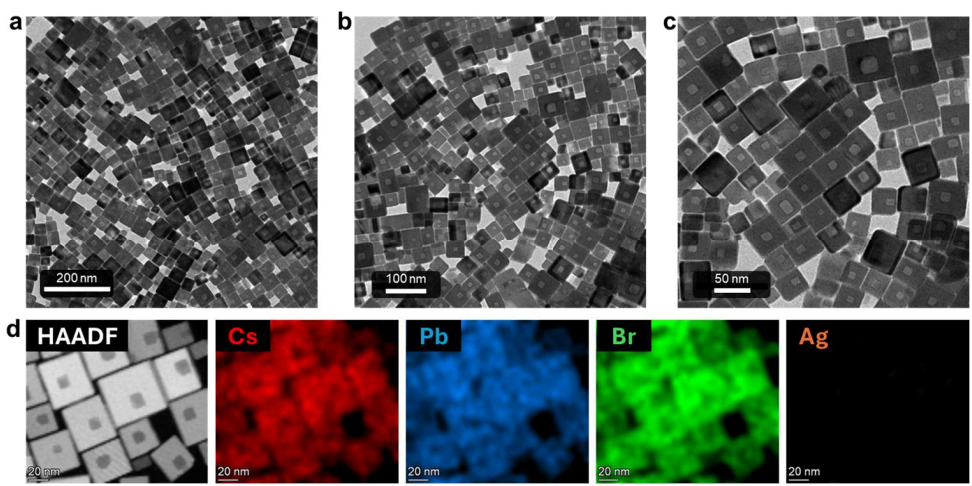

**Figure S15.** (a-c) TEM images and (d) STEM-EDX maps of hollow $CsPbBr_3$ NCs.

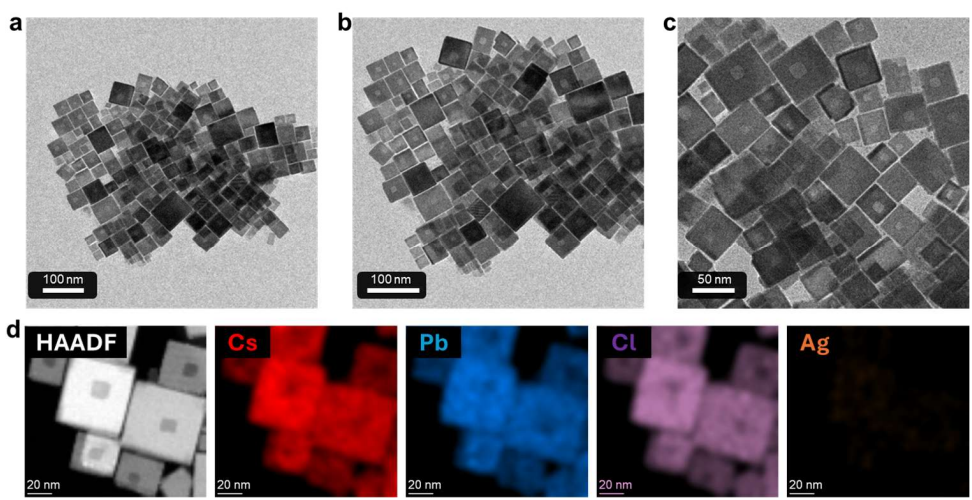

**Figure S16.** (a-c) TEM images and (d) STEM-EDX maps of hollow $CsPbCl_3$ NCs.

ADDITIONAL REFERENCES

(1) Toso, S.; Dardzinski, D.; Manna, L.; Marom, N. Structure Prediction of Ionic Epitaxial Interfaces with Ogre Demonstrated for Colloidal Heterostructures of Lead Halide Perovskites. ACS Nano **2025**, *19* (15), 5326–5341.